\documentclass[aps,twocolumn,showpacs,superscriptaddress,groupedaddress,nofootinbib]{revtex4} 
\usepackage{graphicx}
\usepackage[sort&compress]{natbib}
\usepackage{subfigure}
\usepackage{amsmath}
\usepackage{amssymb}
\usepackage{amsfonts}
\usepackage{rotating}
\usepackage{cancel}
\usepackage{lipsum}
\usepackage{mathtools}
\usepackage{color}
\usepackage{bbm}
\usepackage{dsfont}
\usepackage{bbold}
\usepackage{multirow}
\usepackage{ulem}
\usepackage{slashed}

\begin{document}

\title{Quark mass dependence of the low-lying charmed mesons at one loop in HH$\chi$PT}
\author{F. Gil-Dom\'inguez}
\email{fernando.gil@ific.uv.es}
\affiliation{Departamento de F\'{\i}sica Te\'orica and IFIC,
Centro Mixto Universidad de Valencia-CSIC, Parc Científic UV, C/ Catedrático José Beltrán, 2, 46980 Paterna, Spain}  

\author{R. Molina}
\email{Raquel.Molina@ific.uv.es}
\affiliation{Departamento de F\'{\i}sica Te\'orica and IFIC,
Centro Mixto Universidad de Valencia-CSIC, Parc Científic UV, C/ Catedrático José Beltrán, 2, 46980 Paterna, Spain}   

\begin{abstract}  
We study the light- and heavy-quark mass dependence of the low-lying charmed mesons in the framework of one-loop HH$\chi$PT. The low-energy constants are determined by analyzing the available lattice data from different LQCD simulations. Model selection tools are implemented to determine the relevant parameters as required by the data with higher precision. Discretization and other effects due to the charm quark mass setting are discussed.
\end{abstract}
\maketitle
\section{Introduction}
The discovery of exotic hadrons in the heavy quark sector, which cannot be accommodated in terms of $q\bar{q}$ mesons or $qqq$ baryons, like those with a tetraquark or pentaquark structure, has manifested the relevance of hadronic loops in order to explain the masses and other properties of many states in the hadron spectrum \cite{guoreview}. 

Even there have been claims for a long time of such effects for some low-lying states. This is the case, for example, for the scalar mesons. An approach for (pseudoscalar) meson - (pseudoscalar) meson scattering based on chiral symmetry and unitarity, the so-called Unitarized Chiral Perturbation Theory (UChPT) \cite{truong, dobado, dobado2, nievesquique2,Garcia-Recio:2003ejq,ollerosetpelaez,nicola} has predicted the properties of $\sigma$, $a_0$, and $f_0$, which are in reasonably good agreement with experiment \cite{ollerscalars,ollerpelaezchiral,kaiser,locher,nievesquique2}. The study of the large $N_c$ behavior concludes that, while the low-lying scalars are non-ordinary mesons, the vector mesons are dominated by the $q\bar{q}$ interaction \cite{pelaezsigma}. However, on the lattice QCD (LQCD) side, both $q\bar{q}$ and $\pi\pi$ operators are needed to obtain the properties of the $\rho$ meson \cite{aoki,gockeler,feng,lang,pelissier,bali,dehua,erben,Wilson,Dudek,Bulava,Fengxu, Alexandrou,Fu,Metivet,fischer}. The role of the coupling of the $K\bar{K}$ channel to $\pi\pi$ in the $\rho$-meson mass has been studied in \cite{dehua,hu,hu2,raquelelvira}.% from the analysis of LQCD data with UChPT, concluding that this coupling has a larger effect for lower than physical strange quark masses \cite{raquelelvira}. %However, typically, the rho meson mass obtained from $N_f=2$ lattice simulations  \cite{aoki,gockeler,feng,lang,pelissier,bali,dehua,erben} is lower than that found in $N_f=2+1$ lattice simulations \cite{Wilson,Dudek,Bulava,Fengxu, Alexandrou,Fu,Metivet}. In \cite{dehua,hu,hu2,raquelelvira} it is suggested that the coupling of $\pi\pi$ to $K\bar{K}$ might provide a reason for that. There is  controversy since a recent $N_f=2$ lattice simulation \cite{fischer} obtains a rho meson mass compatible with experiment, so that these effects could be due to lattice artefacts. However, the decay width obtained in \cite{fischer} is too large, $\sim 180(6)$ MeV. Still, the analysis of the $N_f=2+1$ lattice data of \cite{Bulava} with kaon masses in the $2\,m_u+m_s=\mathrm{cte}$ trajectory, which studies the $\pi\pi-K\bar{K}$ interaction in coupled channels within one-loop UChPT, has shown an interesting effect, which is that the $\rho$ becomes lighter when reducing the strange quark mass \cite{raquelelvira}.

In the charm sector, it could be easier to dissentangle ordinary from non-ordinary mesons, since many of the observed particles are narrow states. In fact, a plethora of exotic states for which hadron loops might play a relevant role, especially when these states are close to thresholds, have been discovered in recent years \cite{olsen,chen}. Interestingly, it has been found that, when hadron loops are incorporated into the quark model, mass shifts can be quite large even for low-lying charmonium states \cite{barnes}. 

%Indeed, some of the particles recently discovered in the charm sector have masses suspiciously close to the two meson thresholds, like the $D_{s0}(2317)$, the $X_0(2900)$ (flavour exotic), or the $X(3872)$, or to the meson-baryon thresholds, as the pentaquark states observed by the LHCb.

In the heavy-light meson sector, heavy quark symmetry arises in the heavy-quark limit ($m_Q\to\infty$). This involves both Heavy Quark Spin Symmetry (HQSS) and Heavy Quark Flavor Symmetry (HQFS) \cite{isgur1,isgur2,isgur3,manohar}. Heavy quark spin symmetry emerges in this limit because the color magnetic moment of a heavy quark is proportional to the inverse of its mass, $\mu^c \propto 1/m_Q$, and thus the interaction becomes invariant under the SU(2) spin symmetry, i. e., for each $Q\bar{q}$ state there will be another one degenerate in mass, which can be obtained by a flip of the spin of the heavy quark. The spin-parity of the heavy and light degrees of freedom is then separately conserved. Beyond that, the exact value of the heavy quark mass in the interactions with the light degrees of freedom plays no role, and this gives rise to heavy flavor symmetry. Obviously, these symmetries are broken for finite quark masses and terms that go like $1/m_Q$ could be relevant for the charm quark. 

Thus, in the heavy quark limit, one can use the light degrees of freedom (d.o.f) to classify the heavy-light meson states. The total angular momentum of the light d.o.f are $j_l=L\pm \frac{1}{2}$, being $L$ the angular momentum. In the ground state, two degenerate states with $J^P=\{0,1\}^-$ appear for $L=0$ ($j_l^P=\frac{1}{2}^-$), while for $L=1$ there are a couple of doublets, one with $J^P=\{0,1\}^+$ ($j_l^P=\frac{1}{2}^+$), and another for $J^P=\{1,2\}^+$ ($j_l^P=3/2^+$). In reality, in the PDG the $D$ and $D^*$ mesons correspond to the ground state doublet, while $D^*_0(2300)$ and $D_1(2420)$ could be related to the $J^P=\{0,1\}^+$ doublet, and $D_1(2430)$ and $D^*_2(2460)$ would correspond to the doublet with $J^P=\{1,2\}^+$.

Similarly, we find in the PDG their strange counterparts, the $(D_s,D^*_s)$, $(D_{s0}^*(2317)$, $D_{s1}(2460))$, and $(D_{s1}(2536)$, $D_{s2}(2573))$, which should correspond to the doublets with $J^P=\{0,1\}^-$, $\{0,1\}^+$ and $\{1,2\}^+$ respectively. Regarding these states, the constituent quark model predicts broad states decaying to $D^{(*)}K$ for the $J^P=\{0,1\}^+$ doublet with masses higher in about $100$ MeV than the experimental ones \cite{godfrey,godfrey2,dipierro}. In contrast, the observed states, the $D_{s0}^*(2317)$ and $D_{s1}(2460)$, are very narrow resonances that are placed relatively close to the $DK$ and $DK^*$ thresholds respectively in the spectrum \cite{babards0,cleods1}. For these reasons, molecular \cite{barnesclose,kolomeitsevlutz,faesslergutsche,faesslergutsche2,lutzsoyeur,Albaladejo:2018mhb} and tetraquark explanations \cite{chenghou,terasaki,maianipiccinini,dmitrasinovic} have been proposed. On the other hand, the predictions of the constituent quark model for the $J^P=\{1,2\}^+$ ($j_l^P=3/2^+$) doublet are in agreement with the experiment \cite{godfrey,godfrey2,dipierro}. However, molecular components for the same quantum numbers cannot be discarded. In fact, in \cite{molinabranz}, assuming that the $D^*_2(2573)$ is a $D^*K^*$ molecular state, the recenlty observed $T_{cs}$ and $T_{c\bar{s}}$ states \cite{LHCb1,LHCb2,LHCb3} is predicted, see also \cite{molinaoset,tcsmolina}. 

In general, states with an additional strange quark are in principle expected to be $100$ MeV higher in mass, since $m_s/m_d\simeq 20$ MeV with $m_d \approx 5$ MeV; however, this is not the case for the charmed mesons. As reported by the PDG, the masses of $D^*_0(2300)$ and $D_1(2420)$ are similar to those of their strange partners, $D_{s0}^*(2317)$ and $D_{s1}(2460)$. The situation for the $L=1$ charmed mesons is more complex. Measurements of the masses of the $0^+$ state lie between $2300$ and $2400$ MeV \cite{belleabdecay,babarbdecay,focusexcited}. Theoretically, a two pole structure has been proposed in coupled-channel approaches based on chiral Lagrangians, unitarity and heavy quark symmetry, resulting in two states with masses of around $2100$ MeV and $2400-2450$ MeV \cite{kolomeitsevlutz,guoshen,guohanhart,albaladejosoler,Du:2017zvv}. The existence of a lower mass pole is supported by Lattice QCD, which consistently finds the mass of the $D^*_0(2300)$ well below that of the $D^*_{s0}(2317)$, in contrast to the experiment \cite{moirpeardon,gayerlang}. %In fact, an analysis of recent experimental data of the LHCb on $B\to D\pi\pi$ where chiral symmetry is imposed, suggests that the mass of the $D^*_0(2300)$ should be much lighter. Contrary to these findings, the tetraquark model of Ref. \cite{dmitrasinovic,dmitrasinovic2} predicts that the $0^+$ states, $D^*_0(2300)$ and $D^*_{s0}(2317)$, are degenerate. 

A strongly interacting system at low energies composed of heavy and light mesons with the heavy meson being of $Q\bar{q}$ type can be studied by means of Heavy Hadron Chiral Perturbation Theory (HH$\chi$PT) \cite{wise,burdman,yancheng}. This is an effective field theory based on both chiral and heavy quark symmetries, where a systematic two double expansions in the parameters $Q/\Lambda_\chi$ and $\Lambda_{QCD}/m_Q$ can be made, $Q$ the momentum transfer (soft scale), $Q\sim m_\pi\sim p_\pi$, $m_Q$, is the mass of the heavy quark, and $\Lambda_\chi=4 \pi f\simeq 1$ GeV (hard scale). HH$\chi$PT has been applied to evaluate the ground state heavy meson masses in the heavy quark limit \cite{isgur1,isgur2}. Leading order corrections to these masses due to the finite heavy quark mass have been studied in \cite{rosnerwise,randallsather,dibartolomeo,blokkorner,jenkins,yehlee} for non-zero light quark masses. The $j^P_l=\frac{1}{2}^+$ doublet has also been introduced as an explicit degree of freedom in \cite{mehenspringer,ananthanarayan,alhakami} where the masses of both doublets were studied including one-loop corrections.

Given the large number of parameters in one-loop HH$\chi$PT it has not been possible to yet pin down precisely the parameters of the theory. Because of the recent progress in lattice QCD it is worth coming back to this issue. In this article we study the quark mass dependence of the ground state charmed mesons, $D$, $D^*$, $D_s$ and $D^*_s$ within one-loop HH$\chi$PT \cite{jenkins} by analyzing the available lattice data. There are several reasons to do this: (1) to test the predictive power of HH$\chi$PT with the available lattice data, (2) to better understand the significance of one-loop corrections involving light mesons for the ground state charmed mesons, (3) to provide an overview of the trends of charmed mesons in different lattice QCD simulations, and (4), since many exotic states have been observed in the recent years \cite{guoreview,babards0,cleods1,LHCb1,LHCb2, LHCbnature}, the analysis done here can be used in the future to investigate the quark mass dependence of these new states. This could help to discriminate between different pictures (tetraquark, molecule ...). 

We investigate both, the light and heavy quark (near the physical point) mass dependence of the ground state charmed mesons, $D_{(s)}$ and $D_{(s)}^*$. We investigate which parameters of one-loop HH$\chi$PT are more relevant as required by the LQCD data implementing model selection tools, as the Least Absolute Shrinkage and Selection Operator (LASSO) method in combination with cross validation and information criteria~\cite{ISL,orilasso,landay}. This can be possible now since there are sufficient lattice simulations to better constraint the parameters of HH$\chi$PT. For this analysis, we have taken as input the lattice data on the ground state charmed mesons of Ref. \cite{kalinowskiwagner,EuropeanTwistedMassa} (ETMC), the PACS-CS ensembles \cite{mohlerwoloshyn,aokiphys}, the ensembles from Hadron Spectrum Collaboration (HSC) \cite{cheungohara,cheungthomas}, the data of Table 1 of \cite{prelovsekpadmanath}, the data from Ref. \cite{brunomattia} (CLS), and that of Refs. \cite{balicollins,balicollinsa} (RQCD). Most of these data are collected in Tables tables VIII-XIV of \cite{guoheo}, where a previous lattice data analysis using the chiral lagrangians developed in \cite{lutzsoyeur,kolomeitsevlutz} was performed. In addition, we incorporate the latest studies of charmed meson masses from MILC in $N_f=2+1+1$ \cite{milc4,bddecays}, which have been done for different quark masses and lattice spacings.

%Recently, flavour exotic states have been observed in the LHC which are very close to the $D^*\bar{K}^*$ threshold, and should have a minimum quark content of $ud\bar{c}\bar{s}$. These states were predicted earlier in the framework of the Local Hidden Gauge approach. See also related works [Geng38,39]. Compact tetraquark explanations have also being proposed in [34,35pdg].
\section{Formalism}
\subsection*{Heavy meson masses to one-loop order in HH$\chi$PT}
In HH$\chi$PT ~\cite{wise,burdman,yancheng} it is customary to introduce a single field, $H_l$, with $l=1,2,3$, to refer to $u,d,s$ respectively, including both the pseudoscalar and vector meson fields defined as, $P_l^{(*)}=(D^{(*)0},D^{(*)+},D_s^{(*)})$, for the charmed mesons (and similarly for the bottom mesons),
\begin{equation}
 H^{(Q)}_l=\frac{(1+\slashed{v})}{2}(P^{*(Q)}_{l\mu}\gamma^\mu-P^{(Q)}_l\gamma_5)\ .
\end{equation}
with $Q=c,b$. Similarly, the conjugated meson field is defined as,
\begin{equation}
 \bar{H}^{(Q)}_l=\gamma^0 H_l^{(Q)\dagger}\gamma^0=(P^{*(Q)\dagger}_{l\mu}\gamma^\mu+P^{(Q)\dagger}_l\gamma_5)\frac{(1+\slashed{v})}{2}\ .
\end{equation}
These heavy meson fields transforms as a doublet under heavy quark spin symmetry and like an anti-triplet under $SU(3)_V$ flavor symmetry, i.e.,
\begin{equation}
 H_l^{(Q)}\to S_Q(H^{(Q)}U^\dagger)_l\end{equation}
 and
 \begin{equation}
 \bar{H}^{(Q)}_l\to (U\bar{H}^{(Q)})_lS^\dagger_Q\ ,
\end{equation}
for the heavy meson field and its conjugate, respectively. A velocity dependent field can also be defined as~\cite{georgi},
\begin{equation}
 H^{(Q)}_v=\sqrt{m_H}e^{i\,m_H}\slashed{v}v_\mu x^\mu H^{(Q)}(x)\ ,\label{eq:mh}
\end{equation}
which satisfies $\slashed{v}H_v=H_v$ and $H_v\slashed{v}=-H_v$. The mass $m_H$ in Eq.~(\ref{eq:mh}) can be expanded in terms of powers of $1/m_Q$ starting in $m_H=m_Q$. The chiral lagrangian involving heavy meson fields defined as in Eq.~(\ref{eq:mh}) and pseudoscalar fields, has a systematic expansion in $1/m_H$, derivative and loop expansion. At leading order, the chiral lagrangian reads \cite{jenkins},
%\begin{widetext}
\begin{align}
 {\cal L}^0=&-i\,\mathrm{Tr}\bar{H}_v(v\cdot \partial)H_v+i\,\mathrm{Tr}\bar{H}_vH_v(v\cdot V)\nonumber\\&+2g\,\mathrm{Tr}\bar{H}_vH_v(S_{lv}\cdot A)\label{eq:lag0}
\end{align}
%\end{widetext}
where $S_{lv}$ stands for the velocity dependent spin operator $S^\mu_{lv}$ \cite{bchpt}, which satisfies $v\cdot S_{lv}=0$, and in the rest frame of the heavy meson, $v^\mu=(1,0)$, this is, $S^\mu_{l}=(0,\vec{\sigma}/2)$. 
The coupling $g$ can be fixed to obtain the experimental radiative and pion decay widths of the charmed vector mesons, we set $g^2=0.55$. 
The pseudoscalar meson fields appear in the combinations,
\begin{equation}
V^\mu=\frac{1}{2}(\xi\partial^\mu\xi^\dagger+\xi^\dagger \partial^\mu \xi)\ , \quad
 A^\mu=\frac{i}{2}(\xi\partial^\mu\xi^\dagger-\xi^\dagger \partial^\mu \xi)\ ,
\end{equation}
%\begin{eqnarray}
% &&V^\mu=\frac{1}{2}(\xi\partial^\mu\xi^\dagger+\xi^\dagger \partial^\mu \xi)\ ,\nonumber\\
% &&A^\mu=\frac{i}{2}(\xi\partial^\mu\xi^\dagger-\xi^\dagger \partial^\mu \xi)\ ,
%\end{eqnarray}
where $\xi=\sqrt{U}$, $U=e^{i\phi/\sqrt{2}f}$, and
\begin{align}
  \phi=&
  \left({\begin{array}{ccc}
        \tfrac{\pi^0}{\sqrt 2}+\tfrac{1}{\sqrt{6}}\eta & \pi^+ & K^+ \\
        \pi^- & -\tfrac{\pi^0}{\sqrt{2}}+\tfrac{1}{\sqrt{6}}\eta & K^0 \\
        K^- & \bar K^0 & -\tfrac{2}{\sqrt{6}}\eta\\
      \end{array}}\right). \nonumber
\end{align}
For the one-loop calculation of the heavy meson masses, terms with up to two insertions of the light quark matrix $M=\mathrm{diag}(m_u,m_d,m_s)$, are needed because counterterms at one-loop order are proportional to two powers of the light quark masses. Thus, the relevant lagrangian needed for the one-loop calculation is given by \cite{jenkins},
\begin{widetext}
 \begin{align}
  {\cal L}^{\slashed{s}}_v=&\,\sigma \,\mathrm{Tr}M_\xi\mathrm{Tr}\bar{H}_vH_v+a\,\mathrm{Tr}\bar{H}_vH_vM_\xi+b\,\mathrm{Tr}\bar{H}_vH_v M_\xi M_\xi+c\,\mathrm{Tr}M_\xi \mathrm{Tr}\bar{H}_vH_vM_\xi+d\,\mathrm{Tr} M_\xi M_\xi \mathrm{Tr}\bar{H}_vH_v\nonumber\\&-\frac{\Delta}{8}\mathrm{Tr}\bar{H}_v\sigma^{\mu \nu}H_v\sigma_{\mu \nu}-\frac{\Delta^{(a)}}{8}\mathrm{Tr}\bar{H}_v\sigma^{\mu\nu}H_v\sigma_{\mu\nu}M_\xi-\frac{\Delta^{(\sigma)}}{8}\mathrm{Tr}M_\xi \mathrm{Tr}\bar{H}_v\sigma^{\mu\nu}H_v\sigma_{\mu\nu}\ ,\label{eq:1l}
 \end{align}
 \end{widetext}
 %\begin{align}
 % {\cal L}^{\slashed{s}}_v=&\,\sigma \,\mathrm{Tr}M_\xi\mathrm{Tr}\bar{H}_vH_v+a\,\mathrm{Tr}\bar{H}_vH_vM_\xi+b\,\mathrm{Tr}\bar{H}_vH_v M_\xi M_\xi \nonumber \\ &+c\,\mathrm{Tr}M_\xi \mathrm{Tr}\bar{H}_vH_vM_\xi+d\,\mathrm{Tr} M_\xi M_\xi \mathrm{Tr}\bar{H}_vH_v\nonumber\\&-\frac{\Delta}{8}\mathrm{Tr}\bar{H}_v\sigma^{\mu \nu}H_v\sigma_{\mu \nu}-\frac{\Delta^{(a)}}{8}\mathrm{Tr}\bar{H}_v\sigma^{\mu\nu}H_v\sigma_{\mu\nu}M_\xi\nonumber\\&-\frac{\Delta^{(\sigma)}}{8}\mathrm{Tr}M_\xi \mathrm{Tr}\bar{H}_v\sigma^{\mu\nu}H_v\sigma_{\mu\nu}\ ,\label{eq:1l}
 %\end{align}
where $M_\xi=\frac{1}{2}(\xi M\xi+\xi^\dagger M\xi^\dagger)$ and $M$ is the (diagonal) light quark mass matrix. Here it is assumed $m=m_u=m_d$. If considering isosping breaking, extra terms should be added, see Eq.~(2.14) of \cite{jenkins}. In general, operators which respect HQSS have an expansion which starts at ${\cal O}(1)$, while the ones violating HQSS, start at ${\cal O}(\frac{1}{m_Q})$. The coefficients $a$ and $\sigma$ are dimensionless. These are functions of $\frac{1}{m_Q}$ starting at ${\cal O}(1)$ and therefore terms with these coefficients respect HQSS. These terms with $a$ and $\sigma$ give rise to the $SU(3)_V$ violating mass splittings among the $P^{(*)}$ mesons and to a singlet contribution to the masses, respectively. Terms accompanying the coefficients $b,c$ and $d$, correspond to chiral lagrangian terms with two insertions of the light quark mass matrix, preserving also heavy quark spin symmetry. The hiperfine splittings of the heavy mesons depend on the coefficients $(\Delta,\Delta^{(\sigma)},\Delta^{(a)})$, which start at ${\cal O}(\frac{1}{m_Q})$, and therefore, terms accompanying these coefficients violate HQSS. If one considers only terms with no insertions of the light quark matrix, one has $M_{P^*_a}-M_{P_a}=\Delta$, which is the hyperfine mass splitting at tree level. Terms with $\Delta^{(\sigma)}$ and $\Delta^{(a)}$ in Eq.~(\ref{eq:1l}) correspond to one insertion of the light quark mass matrix. While the former yields a light quark flavor dependent term, the latter leads to a flavor singlet contribution. Both of them contribute to counterterm coefficients in the $M_{P^*}-M_P$ splitting.

The one-loop masses of heavy mesons can be written in terms of two linear combinations, the spin average term, $\frac{1}{4}(M_{P_a}+3M_{P^*_a})$ and the hyperfine splitting, $M_{P^*}-M_P$, which respects and violate heavy quark spin symmetry, respectively. These, which have been calculated to one-loop in Ref.~\cite{jenkins}, are
\begin{widetext}
\begin{eqnarray}\label{eq:md1}
\frac{1}{4}(M_{P_l}+3M_{P^*_l})&= &\,m_H + \alpha_l-\sum_{X=\pi,K,\eta}\beta_l^{(X)}\frac{M_X^3}{16\pi f^2} +\sum_{X=\pi,K,\eta}\left(\gamma_l^{(X)}-\lambda_l^{(X)}\alpha_l\right)\frac{M_X^2}{16\pi^2 f^2}\log\left(M_X^2/\mu^2\right)+c_l\\\label{eq:md2}
M_{P^*_l}-M_{P_l}&= &\,\Delta+\sum_{X=\pi,K,\eta}\left(\gamma_l^{(X)}-\lambda_l^{(X)}\Delta\right)\frac{M_X^2}{16\pi^2 f^2}\log\left(M_X^2/\mu^2\right)+\delta c_l\ 
\end{eqnarray}
\end{widetext}
where the coefficients $\alpha_l$, $\beta_l^{(X)}$, $(\gamma_l^{(X)}-\lambda_l^{(X)}\alpha_l)$, $c_l$, $(\gamma_l^{(X)}-\lambda_l^{(X)}\Delta)$ and $\delta c_l$ are given in the appendix A. The coefficients $\alpha_l$, $(\gamma_l^{(X)}-\lambda_l^{(X)}\alpha_l)$, $c_l$, and $\delta c_l$ are proportional to powers of the light quark masses while $\beta_l^{(X)}$ and $(\gamma_l^{(X)}-\lambda_l^{(X)}\Delta)$ accompany terms proportional to $M_X$, the mass of the $X$ pseudoscalar meson, $\pi$, $K$ or $\eta$. Thus, $m_H$ and $\Delta$ can be interpreted as the spin-average mass and hyperfine splitting of the heavy mesons in the SU(3) chiral limit. The scale $\mu$ in Eqs.~(\ref{eq:md1}) and (\ref{eq:md2}) is fixed to $770~$MeV. 

\section{Fitting procedure with the LASSO method}
We collect the available lattice data on low-lying charmed mesons included here in Tables~\ref{tab:ETMCJPsi}-\ref{tab:rqcd}, %\ref{tab:ETMCetac}, \ref{tab:pacs}, \ref{tab:hsc}, \ref{tab:cls}, \ref{tab:prel} and \ref{tab:rqcd} 
of the Appendix \ref{app:data}. These include,
\begin{itemize}
 \item Data from Ref. \cite{kalinowskiwagner}, which uses Wilson twisted mass lattice QCD with 2+1+1 dynamical quark flavors for pion masses in the range of $225-445$ MeV, $L\sim 2-3$ fm and three lattice spacings $a=0.0619, 0.0815$ and $0.0885$ fm \cite{EuropeanTwistedMassa} [ETMC] (Tables~\ref{tab:ETMCJPsi} and \ref{tab:ETMCetac}).
 \item The PACS ensembles with $N_f=2+1$ flavor Clover-Wilson configurations \cite{mohlerwoloshyn} for pion masses in the range $m_\pi\sim 150-410$ MeV, $L\simeq 2.9$ fm and a lattice spacing $a=0.0907$ MeV \cite{aokiphys} [PACS-CS] (Table \ref{tab:pacs}).
 \item The ensembles from Hadron Spectrum Collaboration for $N_f = 2 + 1$ flavors of dynamical
quarks with an anisotropic (clover) action \cite{cheungohara}, with $a_s\sim 0.12$ fm, $L\sim 1.9-3.8$ fm and $m_\pi\simeq 240, 390$ MeV. See also Table 2 of  \cite{cheungthomas} [HSC] (Table \ref{tab:hsc}).
\item The data of Table 1 of \cite{prelovsekpadmanath} for a $N_f = 2 + 1$ simulation with Wilson dynamical fermions provided by the Coordinated Lattice Simulations (CLS) with a pion mass of $m_\pi=280$ MeV, $L\simeq 2.1$ fm and a lattice spacing of $a=0.08636$ fm \cite{brunomattia} [CLS] (Table \ref{tab:cls}).
\item Data collected in \cite{langmohler} for $N_F=2$ dynamical light quarks with improved Wilson fermions, a lattice spacing $a=0.1239$, $L\simeq 2$ fm and $m_\pi\simeq 266$ MeV [Prelovsek et al.] (Table \ref{tab:prel}). 
\item The data of the simulation with $N_f=2$ improved clover Wilson sea quarks collected in Table 1 of \cite{balicollins} for $m_\pi=150,290$ MeV, $L\sim 1.7-4.5$ fm and $a=0.07$ fm \cite{balicollinsa} [RQCD] (Table \ref{tab:rqcd}).
\item The latest studies of charmed meson masses from MILC with highly improved staggered-quark ensembles with four flavors of dynamical quarks  \cite{milc4,bddecays}, which have been done for different quark masses, $m_\pi\sim 130-330$, volumes $L\sim 2.5-6$ fm and lattice spacings ranging from $0.03-0.15$ fm (Table I of \cite{bddecays}) allowing for a continuous extrapolation \cite{Komijani} [MILC]\footnote{We use as input the function in the continuum limit from MILC \cite{Komijani} for the $D$ and $D_s$ masses.}.
\end{itemize}
Some of these data were collected in Tables VIII-XIV of \cite{guoheo}. Indeed, the data of PACS-CS and ETMC used here correspond to Tables VIII, X and XI of \cite{guoheo}, where a previous lattice data analysis using the chiral lagrangians developed in \cite{lutzsoyeur,kolomeitsevlutz} was performed. The data of Tables XII and XIII of \cite{guoheo} based on MILC ensembles of the HPQCD and LHPC collaborations are not included here, because instead we include the latest MILC studies of \cite{milc4,bddecays} which supersede the older data in finest lattice spacing and bigger volumes that yielded a continuous extrapolation \cite{Komijani}.

In this work we will analyze the above data with one-loop order HH$\chi$PT given by Eqs.~(\ref{eq:md1}) and (\ref{eq:md2}), looking for consistency between the different sets of data and trying to find a unique solution of the charmed meson masses that provides both, the heavy and light quark mass dependence of those. At present, this can only be achieved considering data from different collaborations. Regarding this, several aspects need to be discussed. First, we observe that, how the charm quark is taken to the physical point and which value is taken the observable to do so, can explain largely the deviations observed by the different collaborations, concluding that these actually come mostly from different charm quark masses at the so-called ``physical point''. To compare the charm quark mass settings of the LQCD data used here, in Table~\ref{tab:comp} we compare the value of $m_{D_s}$ in physical units for the lightest pion mass and finest lattice spacing of every collaboration. 
\begin{table}[h!]
 \setlength{\tabcolsep}{0.8em}
{\renewcommand{\arraystretch}{1.6}
\centering
\begin{tabular}{|c|c|c|c|c|}
\hline
 Col. & $a$ & $L$ & $m_\pi$ & $m_{D_s}$ \\
 \hline
ETMC1 & 0.0619 & 3.0 & 224 & 1988 \\
 ETMC2 & 0.0619 & 3.0 & 224 & 2001 \\
 \hline
 PACS-CS & 0.0907 & 2.9 & 156 & 1809 \\
 \hline
 HSC & 0.12 & 3.8 & 239 & 1967 \\
 \hline
 CLS & 0.08636 & 2.1 & 280 & 1981 \\
 \hline
 Prelovsek et al.& 0.1239 & 2.0 & 266 & 1657 \\
 \hline
 RQCD & 0.071 & 4.5 & 150 & 1977 \\
 \hline
\end{tabular}}
\caption{Comparison of the charm quark mass settings of the different collaborations through their value of $m_{D_s}$. Note that the physical value is $m_{D_s}=1968.35\pm 0.07$~MeV \cite{pdg}. $L$ and $a$ are given in fm, and the masses in units of MeV.}
\label{tab:comp}
\end{table}
For the value of the lattice spacing given in \cite{cheungthomas} determined through the $\Omega$ baryon mass, the mass of the $D_s$ for HSC is physical for its light pion mass, the one for RQCD and CLS is around $10$~MeV above the physical one, $20-30$~MeV above for ETMC, while being much lower for the PACS-CS, and [Prelovsek et al.] data sets in $\sim 160$ and $\sim 300$~MeV, respectively. 
Second, there might be discretization effects of different size. In general, we consider that the systematic effects due to discretization are larger for the charm quark mass and that these can be absorbed in the parameters $m_H$ and $\Delta$, as follows,
\begin{eqnarray}
 &&m_H=m_H'+r_H {\cal O}(a^k)\label{eq:mha}\\
 &&\Delta=\Delta'+r_\Delta {\cal O}(a^k)\ ,\label{eq:da}
\end{eqnarray}
where the term dependent on the lattice spacing $a$ will be different for every collaboration depending on the action used, and $m_H'$, $\Delta'$, are just the limit in the continuum of the tree level parameters. Here, we focus on fitting the $m_H$ and $\Delta$ parameters, introducing a new couple of parameters for every different lattice spacing, collaboration, and charm quark mass determination. There are two collaborations which perform simulations at different lattice spacings, MILC and ETMC. For MILC we perform a refit of the continuous extrapolation of \cite{Komijani} that already takes into account the possible systematic effects of the staggered fermions, while for ETMC we analyze the data first and then study the discretization effects a posteriori making ${\cal O}(a^k)\to a^2$ in Eqs.~(\ref{eq:mha}) and (\ref{eq:da}).

In the next section, first we perform individual fits for every collaboration. Then, we make an attempt to perform a {\textit{global}} fit of data combining the different data sets. When doing so, we realize that the inclusion of the PACS-CS data set increases tremendously the $\chi^2$. For this reason we do not include these data in the global fit. In the case of MILC, we have a continuous function which is possible to fit in combination with the rest of data (we take $10$ data points for the $m_D$ mass of this function spread equally in the $m_\pi$ range of the lattice data, Table I of \cite{bddecays}). Since the discretization effects are embedded in the different $m_H$ and $\Delta$ parameters, there is no reason of concern to mix these various data sets, and indeed, as we will see, we obtain a reasonably good value of the $\chi^2_{\mathrm{d.o.f}}$. Thus, for the {\it global} fit, we have, $7+2n$ parameters, with $n=12$, equals to $31$ parameters, $40$ data points for $m_{D_{(s)}}$ and $30$ data for $m_{D_{(s)}^*}$\footnote{In the case of RQCD (Table \ref{tab:rqcd}) there are no data for $m_{D_s^{(*)}}$.}, and the two equations Eqs.~(\ref{eq:md1}) and (\ref{eq:md2}). Apart from the various $(m_H,\Delta)$ parameters introduced, the other parameters to be fitted in Eqs.~(\ref{eq:md1}) and (\ref{eq:md2}) are, ($\frac{\sigma m_\pi}{B_0}$, $\frac{a m_\pi}{B_0}$, $\frac{b m^3_\pi}{B_0^2}$, $\frac{c m^3_\pi}{B_0^2}$, $\frac{d m^3_\pi}{B_0^2}$, $\frac{\Delta^{(\sigma)} m_\pi}{B_0}$, $\frac{\Delta^{(a)} m_\pi}{B_0}$, where $m_ \pi$ is the physical pion mass, and it is introduced to redefine the parameters in terms of dimensionless quantities. 

Given that the size of the data sample is not very large compared to the number of parameters, one could easily have problems with overfitting the data. In order to avoid that, we follow the procedure of Ref.~\cite{landay}, where the Least Absolute Shrinkage and Selection Operator (LASSO) method in combination with cross validation~\cite{ISL,orilasso} and information criteria were applied for the analysis of photoproduction data with a set of models with different number of parameters. With this method, only those models which contain an optimal number of parameters able to describe the data with enough accuracy are selected. To implement the LASSO, a penalty term is added to the $\chi^2$
\begin{eqnarray}
   && \chi^2(\vec{p},\vec{q}) = \sum_{i}^{N} \frac{\left(y_i - f_i(\vec{p},\vec{q}) \right)^2}{\sigma_{y_i}^2} +P,\nonumber\\&&P=\frac{\lambda}{10} \sum_{i=1}^{7} | p_i|\ .\label{eq:lasso}
\end{eqnarray}
In order to select the best ``$\lambda$'' parameter, we proceed as follows: in every iteration we divide randomly the original data set into three subsets. We keep two subsets as the training data set and one subset as the validation data set. These sets will provide the $\chi^2$ of the training and validation sets, $\chi^2_T$, $\chi^2_V$, respectively. As explained in Refs.~\cite{ISL,orilasso,landay}, while $\chi^2_T$ is a monotonously increasing function of $\lambda$, $\chi^2_V$ has a sweet spot in which the data are not underfitted (large $\lambda$) and not overfitted (small $\lambda$). The fluctuations given by the random selection of the validation set provide the error of this minimum, and also the one of the $\lambda$ value, which is chosen as the $\lambda_\mathrm{opt}>\lambda_\mathrm{min}$, compatible within errors with the minimum of $\chi^2_V(\lambda_\mathrm{min})$.

The result for the $\lambda_\mathrm{opt}$ from cross validation is compared with the one obtained from three information criteria, which are, AIC, AICc, and BIC, defined as,

\begin{eqnarray}
&&\text{AIC} = 2k -2\log(L) = 2k + \chi^2 \ ,  \nonumber\\
&&\text{AICc} = \text{AIC} + \frac{2k(k+1)}{n-k-1} \ , \nonumber \\
&&\text{BIC} = k\log(n) -2\log(L)= k\log(n) + \chi^2 \nonumber ,
\end{eqnarray}
In the above equations, $k$ is the number of parameters that changes with $\lambda$, 
$n$ is the length of the data, and $L$ is the likelihood. The optimal value of $\lambda$ is given by the respective minimum of the criteria. 
\section{Results}\label{sec:results}

\begin{table*}[htb!]
\setlength{\tabcolsep}{0.4em}
{\renewcommand{\arraystretch}{1.6}
\centering
\begin{tabular}{|c|c|c|c|c|c|c|c|} \hline
Fit & $\frac{\sigma m_\pi}{{B_0}}\cdot 10^{2}$ & $\frac{a m_\pi}{{B_0}}\cdot 10^{2}$ & $\frac{b m^3_\pi}{{B_0}^{2}}\cdot 10^{4}$& $\frac{c m^3_\pi}{{B_0}^{2}}\cdot 10^{4}$ & $\frac{d m^3_\pi}{{B_0}^{2}}\cdot 10^{4}$ &  $\frac{\Delta^{(\sigma)}m_\pi}{{B_0}}\cdot 10^{3}$& $\frac{\Delta^{(a)}m_\pi}{{B_0}}\cdot 10^{3}$\\ \hline

ETMC$_{\textrm{(I)}}$ & $1.23 \pm 0.11$ & $2.87 \pm 0.18$ & $-1.45 \pm 0.55$ & $17.11\pm 0.53$ & $8.90\pm 0.47$ & $7.97\pm 4.73$ & $-2.98\pm 3.13$ \\ \hline
ETMC$_{\textrm{(II)}}$ & $1.41 \pm 0.01$ & $2.79\pm 1.63$ & - & $16.18 \pm 0.63$ & $8.51\pm 0.71$ & $8.11\pm 5.31$ & $-2.99 \pm 3.09$ \\ \hline

MILC$_{\textrm{(I)}}$ & $4.14 \pm 30.36$ & $4.14 \pm 2.48$ & $18.40 \pm 283.83$ & $5.26\pm 144.54$ & $4.10\pm 18.60$ & $5.52\pm 35.19$ & $49.68\pm 65.96$ \\ \hline
MILC$_{\textrm{(II)}}$ & $2.40 \pm 0.55$ & $3.31\pm 1.01$ & - & $14.75 \pm 4.36$ & - & - & - \\ \hline

PACS$_{\textrm{(I)}}$ & $1.24 \pm 0.33$ & $3.61 \pm 0.28$ & $-6.25 \pm 4.76$ & $18.47\pm 3.26$ & $9.28\pm 1.94$ & $4.97\pm 0.22$ & $-1.05\pm 0.61$ \\ \hline
PACS$_{\textrm{(II)}}$ & $1.68 \pm 0.03$ & $3.26\pm 0.04$ & - & $14.20 \pm 0.08$ & $11.83 \pm 0.08$ & $4.95 \pm 0.22$ & $-1.04 \pm 0.61$ \\ \hline

GLOBAL$_{\textrm{(I)}}$ & $2.54 \pm 0.01$ & $3.51 \pm 0.03$ & $1.50 \pm 0.13$ & $11.75\pm 0.11$ & $6.14\pm 0.08$ & $3.73\pm 0.29$ & $-1.97\pm 0.65$ \\ \hline
GLOBAL$_{\textrm{(II)}}$ & $1.61 \pm 0.01$ & $3.56\pm 0.03$ & - & $12.83 \pm 0.13$ & $5.96 \pm 0.08$ & $3.53 \pm 0.28$ & $-2.03 \pm 0.63$ \\ \hline
\end{tabular}}
\caption{Collection of parameters from different fits. The subindex (I) means "before LASSO" and (II) means "after LASSO".}
\label{tab:pars}
\end{table*}

In this section, we first present the results obtained from individual fits for data sets larger enough to be fitted by means Eqs.~(\ref{eq:md1}),~(\ref{eq:md2}) (see also Eqs.~
(\ref{eq:par1}),~(\ref{eq:par2}),~(\ref{eq:par3})). These data sets are: ETMC, PACS, and MILC. Second, we show the result of a {\it global} fit that also includes the rest of the data sets, HQS, CLS, [Prelovsek et al.] \cite{langmohler}, with few data points for heavy meson masses.

All fits are carried out with quantities in units of the lattice spacing ($am_\pi$, $am_D$ etc.). As commented in the previous section, we assume that discretization effects are much larger for the charm quark mass and absorb the overall dependence of the charmed meson masses with the lattice spacing in the $m_H$ and $\Delta$ parameters (we introduce one pair of ($m_H$, $\Delta$) for every lattice spacing). For data sets with more than one lattice spacing, ETMC data, we also analyze this discretization effect after the fit. We will compare this result at $a=0$ with that of MILC (extrapolated data to the continuum) and the result of the global fit.

\subsection{ETMC}
\begin{figure}[h!]
   \begin{minipage}{0.461\textwidth}
     \centering
     \includegraphics[width=1.0\linewidth]{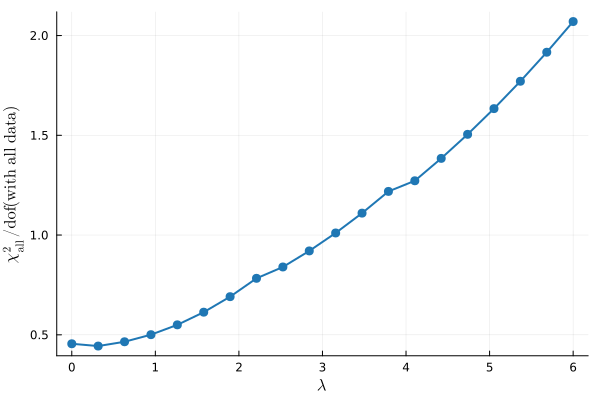}
   \end{minipage}
   \begin{minipage}{0.461\textwidth}
     \centering
     \includegraphics[width=1.0\linewidth]{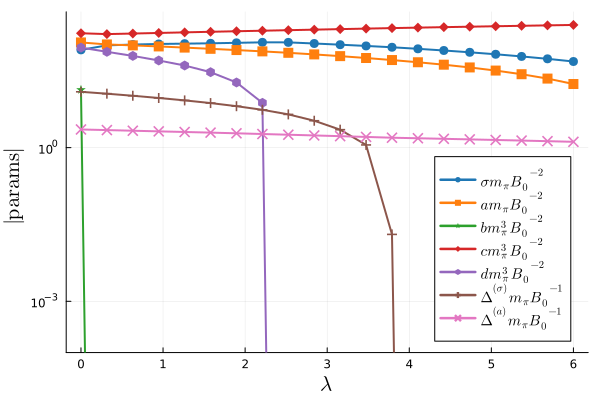}
   \end{minipage}
   \begin{minipage}{0.461\textwidth}
     \centering
     \includegraphics[width=1.0\linewidth]{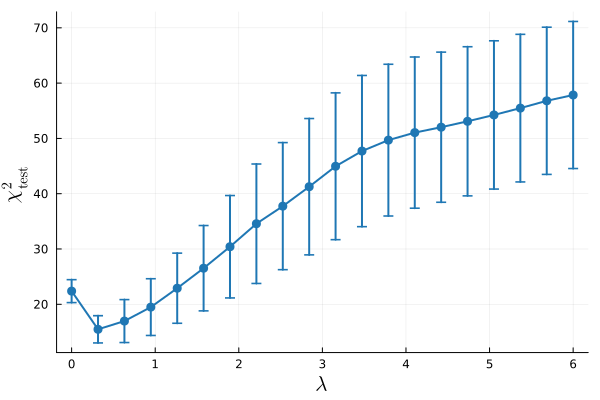}
   \end{minipage} 
   \begin{minipage}{0.461\textwidth}
     \centering
     \includegraphics[width=1.0\linewidth]{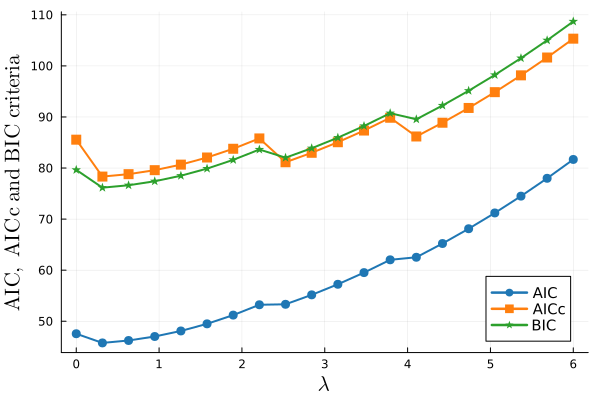}
   \end{minipage}
   \caption{LASSO method plots for ETMC data analysis.}
   \label{fig:LASSOETMC}
\end{figure}

The results after LASSO are collected in Fig.~\ref{fig:LASSOETMC}.

%\cm{@Fer incluye los resultados de ETMC pero solo 4 gráficas en el siguiente orden: $\chi^2_{dof}$ all data arriba-izq, parametros, abajo-izq, $\chi^2_{test}$ arriba-derecha, criterios informacion abajo-derecha. Nota: Manten los colores de las lineas en los siguientes graficos para otras colaboraciones, pon los label y referencias a las figuras correctamente}.

As can be seen in Fig.~\ref{fig:LASSOETMC}, in the first figure, from top to bottom, the $\chi^2_{\mathrm{dof}}$ is a monotonously increasing function with $\lambda$. Fig.~\ref{fig:LASSOETMC} (second) shows the relevance of the parameters according to the lattice data. In this figure it is clear that the data do not constrain well the parameter $b$, which according to the analysis with LASSO, it is a superfluous degree of freedom. Next, $d$ and $\Delta_\sigma$ are also less relevant for the data. The $b$ and $d$ parameters correspond to chiral lagrangian terms with two insertions of the light quark mass matrix preserving HQSS, while $\Delta_\sigma$ violates HQSS and provides higher order corrections to the hyperfine splitting at tree level. Notice that parameters of these both types, like $c$ and $\Delta_a$, are, however, required by the data, i. e., not all the parameters of the same order are equally relevant. 

One should have a criteria in order to decide whether one or the three parameters can be removed. We follow the criteria used in \cite{landay}, explained in the previous section, cross validation and information criteria. These are given in the third and fourth panels of Fig.~\ref{fig:LASSOETMC}, respectively. Information criteria are evaluated with all data, while for cross validation, we divide the data sets in three sets for every lattice spacing, $a$, and select $2$ sets for the training set and $1$ for the validation (test) set. As one can see, the information criteria have three minima corresponding to leaving out, one parameter, $b$, which is also the global minima, two parameters, $b$ and $d$, and three parameters, $b$, $d$ and $\Delta_\sigma$, respectively. The global minima coincides with the one from cross validation, in Fig.~\ref{fig:LASSOETMC} (second). Thus, we select this parametrization dropping out parameter $b$ as the most predictive one. 

The parameters obtained before and after the LASSO, that we call Fits I and II, respectively, are given in the two first rows of Tables~\ref{tab:pars} and \ref{tab:ETMCpars}. 
\begin{table}[htb]
\setlength{\tabcolsep}{0.4em}
{\renewcommand{\arraystretch}{1.6}
\centering
\begin{tabular}{|c|c|c|c|c|c|} \hline
\multicolumn{2}{|c|}{Fit} & \multicolumn{2}{|c|}{$m_H$ (MeV)} & \multicolumn{2}{|c|}{$\Delta$ (MeV)} \\ \hline
Col. & $a$ (fm) & (I) & (II) & (I) & (II) \\ \hline
\multirow{3}{*}{ETMC1} & $0.0619$ & $1995 \pm 12$ & $1988 \pm 16$ & $122 \pm 21$ & $121 \pm 24$ \\ \cline{2-6}
& $0.0815$ & $2014 \pm 13$ & $2007 \pm 17$ & $129 \pm 25$ & $128 \pm 27$ \\ \cline{2-6}
& $0.0885$ & $2015 \pm 14$ & $2008 \pm 18$ & $132 \pm 28$ & $131 \pm 29$ \\ \hline
\multirow{3}{*}{ETMC2} & $0.0619$ & $2008 \pm 12$ & $2001 \pm 16$ & $123 \pm 21$ & $123 \pm 23$ \\ \cline{2-6}
& $0.0815$ & $2038 \pm 13$ & $2031 \pm 16$ & $126 \pm 26$ & $126 \pm 27$ \\ \cline{2-6}
& $0.0885$ & $2047 \pm 14$ & $2041 \pm 18$ & $131 \pm 25$ & $130 \pm 27$ \\ \hline
ETMC & 0 & $1972$ & $1965$ & $114$ & $114 $ \\ \hline
\end{tabular}}
\caption{Tree level parameters of the ETMC fits. The remaining parameters are shown in the first two rows of Table~\ref{tab:pars}.}
\label{tab:ETMCpars}
\end{table}
We see that the result between the two different fits is very similar and that, in fact, the inclusion of the $b$ parameters, do not affect much the values of the rest of the parameters in this case.

\begin{figure}[h!]
   \begin{minipage}{0.422\textwidth}
     \centering
     \includegraphics[width=1.0\linewidth]{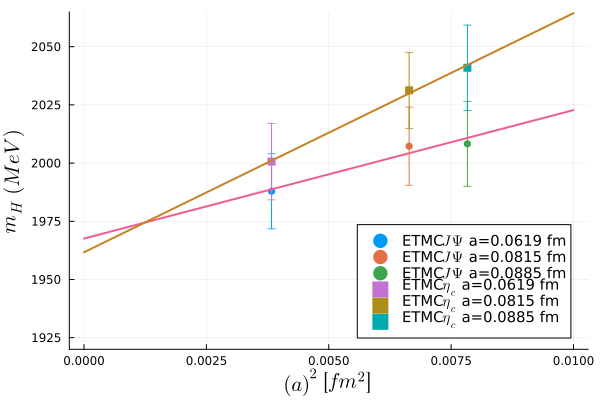}
   \end{minipage}
   \begin{minipage}{0.422\textwidth}
     \centering
     \includegraphics[width=1.0\linewidth]{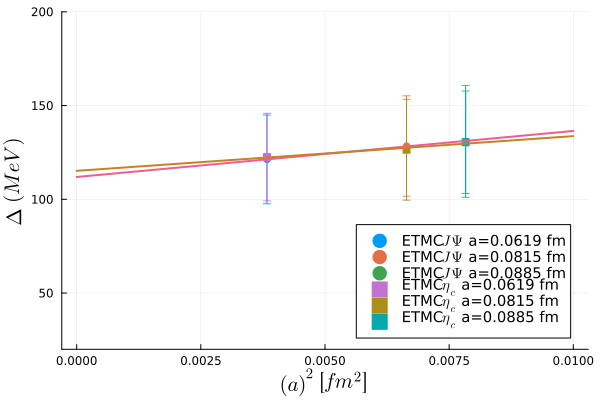}
   \end{minipage} 
   \caption{Continuum extrapolation of $m_H$ and $\Delta$ parameters from the ETMC analysis (Fit II).}
   \label{fig:mHDeltaETMC}
\end{figure}

For these data sets, one can also study the discretization effects. Assuming,
\begin{eqnarray}
 m_H=m_H'+r_Ha^2\nonumber\\
 \Delta=\Delta'+r_\Delta a^2\ ,\label{eq:a}
\end{eqnarray}
with the value of $m_H$ and $\Delta$ for different $a$ obtained in the fit and its errors plotted in Fig.~\ref{fig:mHDeltaETMC}. 
We fit the two different data from fixing the charm quark mass with the $\eta_c$ or $J/\psi$ mass with Eq.~(\ref{eq:a}). We observe that, in these data sets, the difference between fixing the charm quark mass with the $\eta_c$ or $J/\psi$ mass is basically due to discretization effects, since this difference is close to zero for $a\sim 0$, see Fig.~\ref{fig:mHDeltaETMC} (top). For the extrapolation to the continuum of $m_H$, we take the average of the two lines at $a\to 0$. For the hyperfine splitting, we obtain the line in Fig.~\ref{fig:mHDeltaETMC} (bottom). Since this parameter starts its expansion at ${\cal O}(1/m_Q)$, there is no difference between setting the charm quark mass with $\eta_c$ or $J/\psi$, and the error due to discretization effects of this parameter is much smaller, being much more stable when varying the lattice spacing. For $a=0$, we give the values of $m_H$ and $\Delta$ in the last row of Table~\ref{tab:ETMCpars}. 

\begin{figure}[h!]
   \begin{minipage}{0.45\textwidth}
     \centering
     \includegraphics[width=1.0\linewidth]{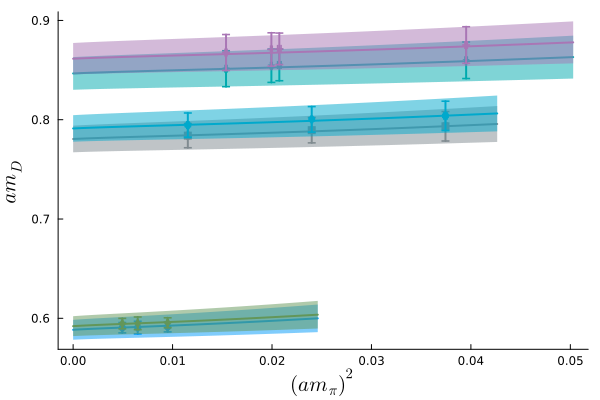}
   \end{minipage}
   \begin{minipage}{0.45\textwidth}
     \centering
     \includegraphics[width=1.0\linewidth]{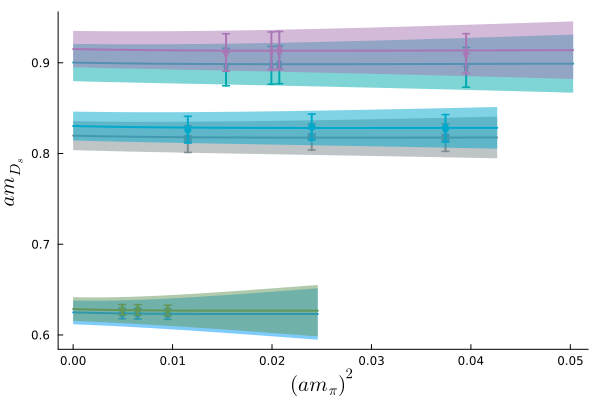}
   \end{minipage} 
   \begin{minipage}{0.45\textwidth}
     \centering
     \includegraphics[width=1.0\linewidth]{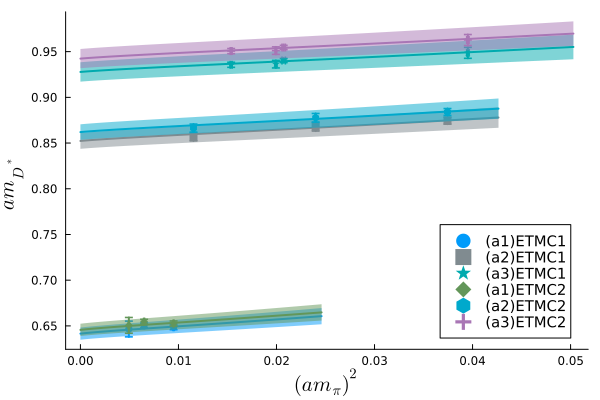}
   \end{minipage}
   \begin{minipage}{0.45\textwidth}
     \centering
     \includegraphics[width=1.0\linewidth]{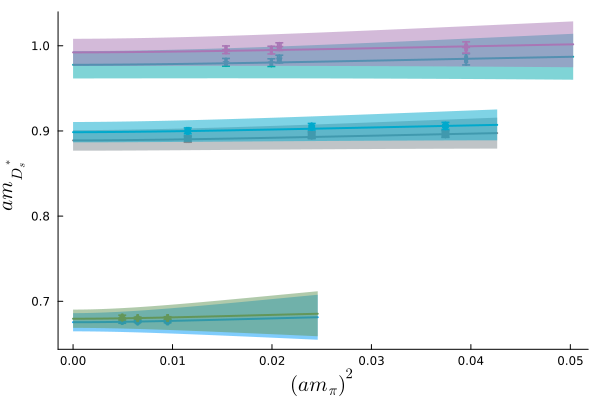}
   \end{minipage}
   \caption{Results of the $D$, $D^*$, $D_s$ and $D^*_s$ meson masses from the ETMC analysis (Fit II).}
   \label{fig:massesETMC}
\end{figure}

We observe that the parameter $m_H$ is indeed very close to the physical value of $(m_D+3m_{D^*})/4=1973$~MeV, however, $\Delta$ differs $\sim20\%$ from the physical hyperfine splitting, $m_{D^*}-m_D=141~$MeV. This motivates us to perform also fits with the expressions of the charmed meson masses at tree level, keeping terms with one insertion of the light quark mass matrix in Eq.~(\ref{eq:1l}). The results are given in the Appendix \ref{app:tree}. While for the one-loop expressions we obtain $\Delta=114$ MeV for $a=0$, for the tree level expressions, Eqs.~(\ref{eq:t1}) and (\ref{eq:t2}), the continuous extrapolation of this parameter that we obtain is $\Delta=129$~MeV. The values of $m_H$ are quite similar for these different fits, $m_H\simeq 1970$. As expected, the one-loop expressions reduce the value of the hyperfine splitting in the chiral limit. We show the results of Fit II in Fig.~\ref{fig:massesETMC}. Eqs.~(\ref{eq:md1}) and (\ref{eq:md2}) provide a very good description of the data. The $\chi^2_{\mathrm{dof}}$ obtained is 0.44. To evaluate the errors, here and in the following, we use automatic differentiation \cite{Ramos:2018vgu} with the Julia ADerrors
package. 

\begin{figure}[h!]
   \begin{minipage}{0.45\textwidth}
     \centering
     \includegraphics[width=1.0\linewidth]{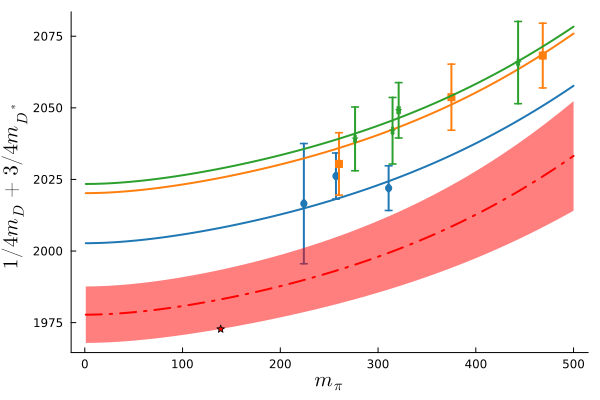}
   \end{minipage}
   \begin{minipage}{0.45\textwidth}
     \centering
     \includegraphics[width=1.0\linewidth]{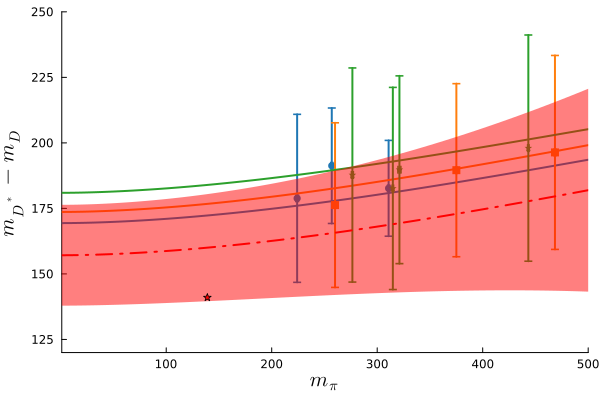}
   \end{minipage} 
   \begin{minipage}{0.45\textwidth}
     \centering
     \includegraphics[width=1.0\linewidth]{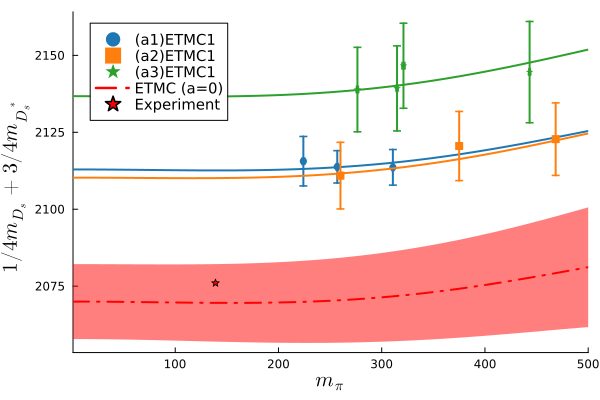}
   \end{minipage}
   \begin{minipage}{0.45\textwidth}
     \centering
     \includegraphics[width=1.0\linewidth]{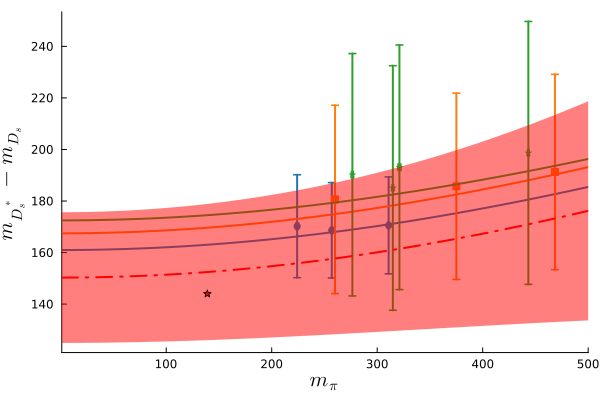}
   \end{minipage}
   \caption{Comparison between ETMC data analysis and the continuum extrapolation obtained.}\label{fig:massesSOLOETMC}
\end{figure}

Finally, we depict in Fig.~\ref{fig:massesSOLOETMC} the extrapolation to the continuum limit of the spin average mass and the hyperfine splitting (red band) in comparison with the ETMC data set. As can be inferred,  discretization effects appear to be of $\sim2-4\%$ in the spin average mass for these data sets. However, for the hyperfine splitting, which suffers from large errors of $\sim 20-30\%$, most of the LQCD data fall inside the error band. In the continuum limit, the LQCD data agree well with the experimental value, shown with a star. 

\newpage
\subsection{MILC}

\begin{figure}[h!]
   \begin{minipage}{0.461\textwidth}
     \centering
     \includegraphics[width=1.0\linewidth]{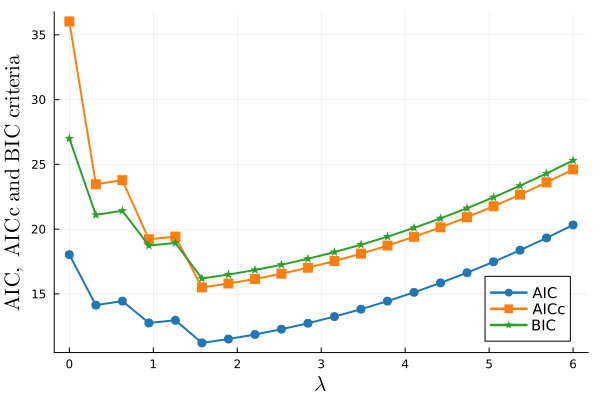}
   \end{minipage}
   \begin{minipage}{0.461\textwidth}
     \centering
     \includegraphics[width=1.0\linewidth]{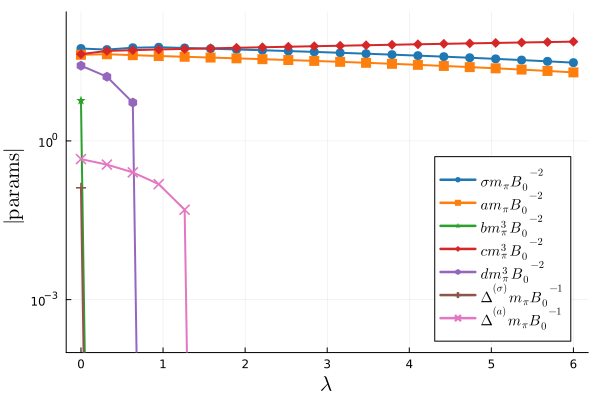}
   \end{minipage} 
   \caption{LASSO method plots for MILC data analysis.}
   \label{fig:LASSOMILC}
\end{figure}

These lattice data are based on the simulations of \cite{milc4,bddecays} for different pion masses and lattice spacings, which made it possible to extrapolate to the continuum the masses of the charmed mesons using EFT's, with the theoretical framework being based on a merger of one-loop HQET and HMrAS$\chi$PT, where a generic lattice spacing dependence, as well as higher order terms, where incorporated \cite{Komijani}. This function has been provided to us and here we conduct a refit of it with Eqs.~(\ref{eq:md1}) and (\ref{eq:md2}). In this case, the $\chi^2_\mathrm{test}$ is a monotonously increasing function and does not show a minimum except for $\lambda=0$. However, the information criteria, shown in Fig.~\ref{fig:LASSOMILC} (top), exhibit a minimum around $\lambda\simeq 2$. Actually, for this value of $\lambda$, four parameters have been dropped out, $b$, $d$, $\Delta_\sigma$ and $\Delta_a$.

\begin{table}[htb]
\setlength{\tabcolsep}{0.4em}
{\renewcommand{\arraystretch}{1.6}
\centering
\begin{tabular}{|c|c|c|c|c|} \hline
\multirow{2}{*}{Collaboration} & \multicolumn{2}{|c|}{$m_H$ (MeV)} & \multicolumn{2}{|c|}{$\Delta$ (MeV)} \\ \cline{2-5}
& (I) & (II) & (I) & (II) \\ \hline
MILC & $1863 \pm 1271$ & $1959 \pm 40$ & $110 $ & $110$ \\ \hline
\end{tabular}}
\caption{Tree level parameters of the MILC fit. The remaining parameters are shown in the first two rows of Table~\ref{tab:pars}.}
\label{tab:MILCpars}
\end{table}
\begin{figure}[h!]
   \begin{minipage}{0.461\textwidth}
     \centering
     \includegraphics[width=1.0\linewidth]{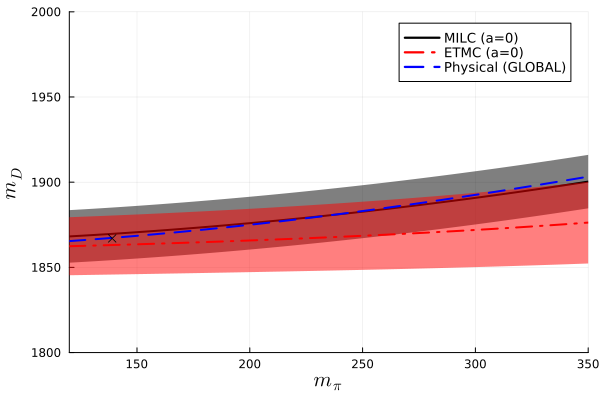}
   \end{minipage}
   \begin{minipage}{0.461\textwidth}
     \centering
     \includegraphics[width=1.0\linewidth]{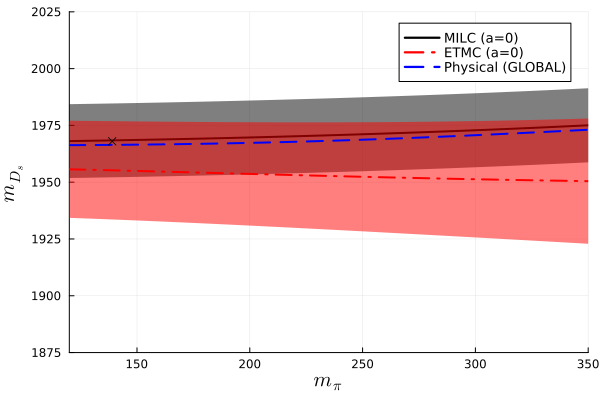}
   \end{minipage} 
   \caption{Continuum extrapolation of charmed meson masses from ETMC (red), MILC data (purple) and physical trajectory extracted from the global analysis (blue).}
   \label{fig:MILCETMCext}
\end{figure}
Note that the first three parameters are also the less relevant parameters for the ETMC data. The values of the parameters, before and after performing the LASSO method, Fits I and II, are given in Tables~\ref{tab:pars} and \ref{tab:MILCpars}. In this case the data and continuous extrapolation only contain the values for the $m_{D_{(s)}}$ masses. Since the masses of the vector counterparts are absent we have to fix the value of $\Delta$. We have tried several values in the range of the quantities obtained in the other fits, and here, fixing $\Delta$ to $110$ MeV seems to provide the best result. Since four parameters are dropped after the LASSO, the values of the parameters in Fits I and II are quite different. The result of the parameters after the LASSO method has much smaller errors, telling that the method is working well. Concretely, in Fit II, the value and error obtained for $m_H$ is quite reasonable and also compatible with the one from the ETMC fit. The $\chi^2_\mathrm{dof}$ obtained in this case is small, 0.04, because of the refit of the function. In Fig.~\ref{fig:MILCETMCext} we depict the charmed meson masses in comparison with the continuous extrapolation of the ETMC data that we did previously and the result of the physical extrapolation for the global fit presented in Sec.~\ref{sub:gl}. We observe that the ETMC data show less dependence of the charmed meson masses with the pion mass than MILC data and the result from the global fit, but the error bands overlap each other and with the experimental charmed meson masses, shown with stars. As could be expected, $m_{D_s}$ does not depend much with the pion mass, since the strange quark mass is close to physical.

We can also compare our result for the charmed meson mass in the SU(3) chiral limit with the one obtained in \cite{milc4,Komijani}. In \cite{milc4} the value of $m_D$ in the SU(3) chiral limit obtained is $1842.7 \pm 5.4$ MeV, while our fit results in  $m_D^{SU(3)}=1877.0 \pm 39.7$ MeV. These two results are consistent.

\subsection{PACS-CS}
\begin{figure}[h!]
   \begin{minipage}{0.461\textwidth}
     \centering
     \includegraphics[width=1.0\linewidth]{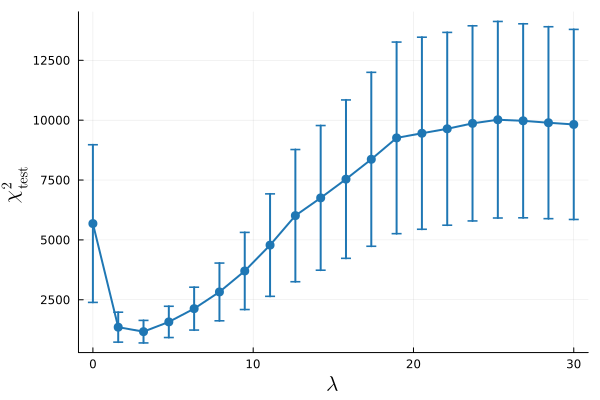}
   \end{minipage}
   \begin{minipage}{0.461\textwidth}
     \centering
     \includegraphics[width=1.0\linewidth]{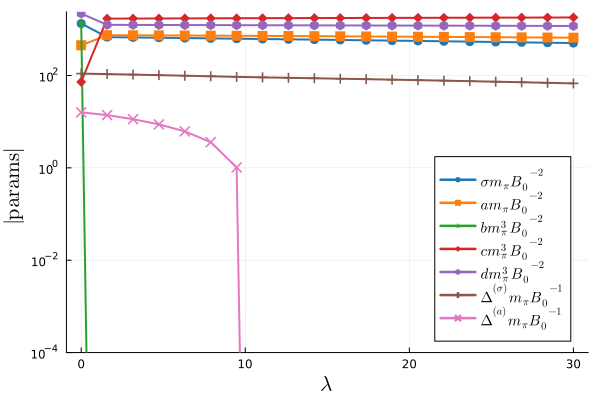}
   \end{minipage} 
   \caption{LASSO method plots for PACS data analysis.}
   \label{fig:LASSOPACS}
\end{figure}

\begin{figure}[h!]
   \begin{minipage}{0.445\textwidth}
     \centering
     \includegraphics[width=1.0\linewidth]{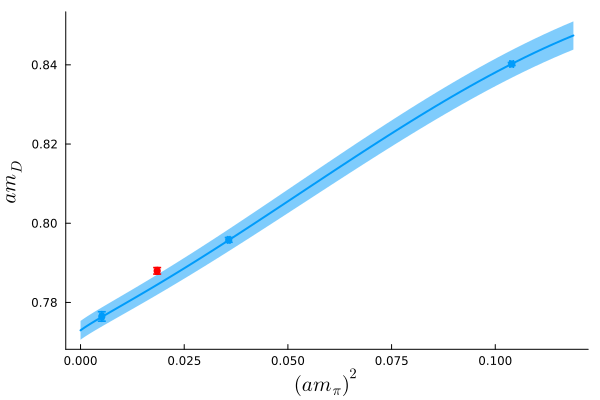}
   \end{minipage}
   \begin{minipage}{0.445\textwidth}
     \centering
     \includegraphics[width=1.0\linewidth]{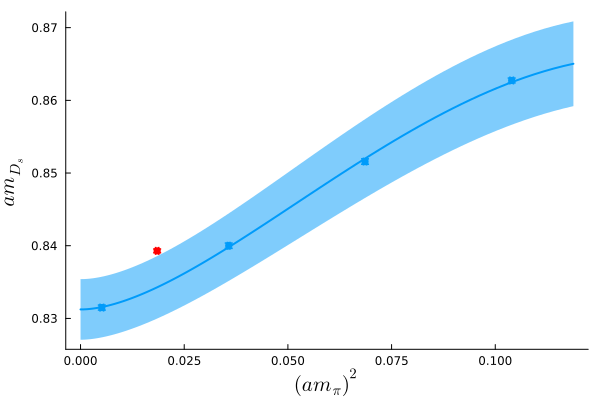}
   \end{minipage} 
   \begin{minipage}{0.445\textwidth}
     \centering
     \includegraphics[width=1.0\linewidth]{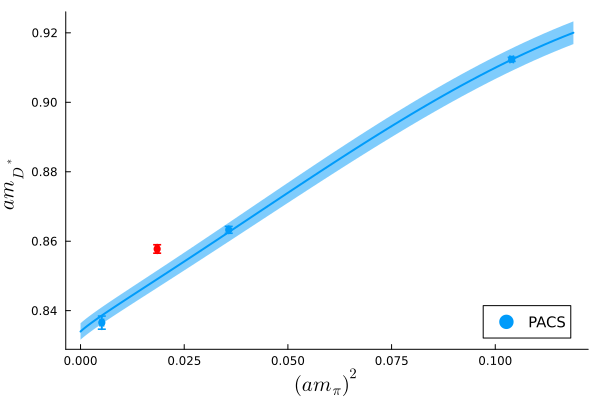}
   \end{minipage}
   \begin{minipage}{0.445\textwidth}
     \centering
     \includegraphics[width=1.0\linewidth]{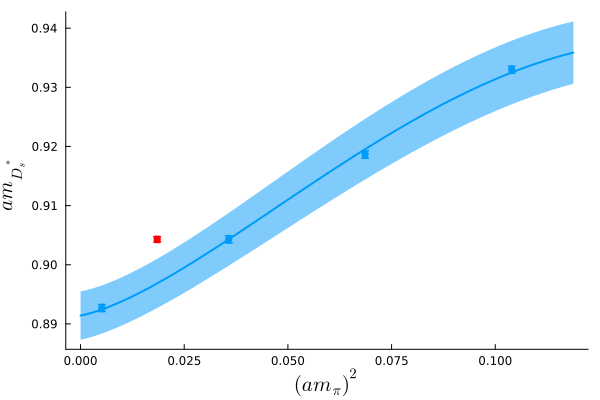}
   \end{minipage}
   \caption{Results of the $D$, $D^*$, $D_s$ and $D^*_s$ meson masses from the PACS analysis (Fit II) after eliminating the fourth pion mass of Table~\ref{tab:pacs}.}
   \label{fig:massesPACS}
\end{figure}

In this case the information criteria did not show a clear minimum, except for $\lambda=0$. However, cross validation gives a clear minimum in $\chi^2_\mathrm{test}$ around $\lambda\simeq 2$, as shown in Fig.~\ref{fig:LASSOPACS} (top), which corresponds to dropping out the $b$ parameter, see Fig.~\ref{fig:LASSOPACS} (bottom).
Nevertheless, the $\chi^2_{\mathrm{dof}}$ obtained is very high (much larger than $1$). This is clearly not consistent with the previous data sets, where very good values of the $\chi^2_{\mathrm{dof}}$ were obtained. We conclude that this lattice data set for the charmed meson masses is in tension with the rest of the lattice data sets. We notice that the high value of the $\chi^2$ for this collaboration is caused by one light pion mass. If we eliminate this point and we refit the data we obtain a $\chi^2_\mathrm{dof}$ of $4.4$. We show the plots of the charmed meson masses as a function of the pion mass in lattice units in Fig.~\ref{fig:massesPACS},
were the conflicting data point is shown in red color. The values of the parameters obtained are given in Tables~\ref{tab:pars} and \ref{tab:PACSpars}. Still, the parameters obtained in Table~\ref{tab:pars} are not so different from the other data sets. However, $m_H$ in Table \ref{tab:PACSpars} is significantly lower, suggesting that the charm quark mass in these simulations is much lower than the experimental, as also suggested in Table~\ref{tab:comp}.
\begin{table}[h!]
\setlength{\tabcolsep}{0.4em}
{\renewcommand{\arraystretch}{1.6}
\centering
\begin{tabular}{|c|c|c|c|c|} \hline
\multirow{2}{*}{Collaboration} & \multicolumn{2}{|c|}{$m_H$ (MeV)} & \multicolumn{2}{|c|}{$\Delta$ (MeV)} \\ \cline{2-5}
 & (I) & (II) & (I) & (II) \\ \hline
PACS & $1767 \pm 47$ & $1706 \pm 3$ & $99 \pm 2$ & $99 \pm 2$ \\ \hline
\end{tabular}}
\caption{Tree level parameters of the PACS fit. The remaining parameters are shown in the first two rows of Table~\ref{tab:pars}.}
\label{tab:PACSpars}
\end{table}

\subsection{Global fit}\label{sub:gl}
\begin{figure}[h!]
   \begin{minipage}{0.453\textwidth}
     \centering
     \includegraphics[width=1.0\linewidth]{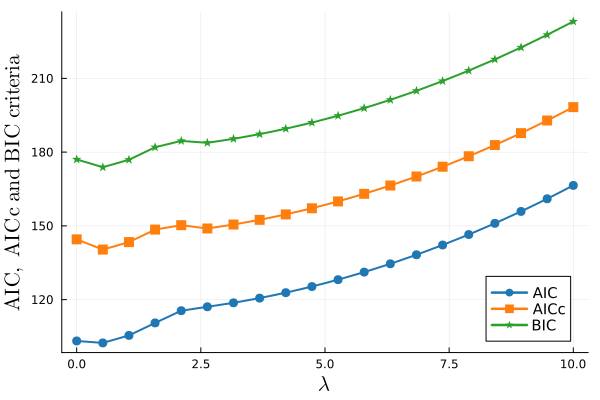}
   \end{minipage}
   \begin{minipage}{0.453\textwidth}
     \centering
     \includegraphics[width=1.0\linewidth]{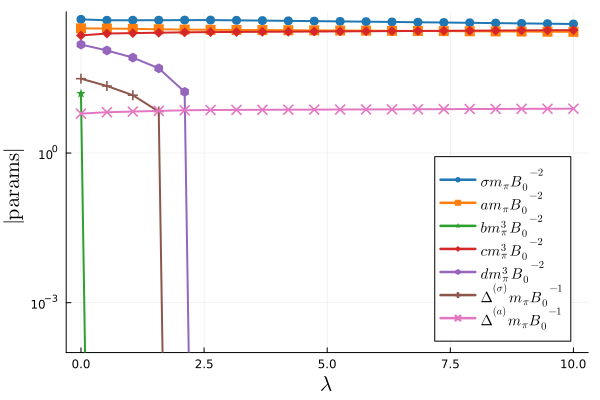}
   \end{minipage}
   \caption{LASSO method plots for the global analysis.}
   \label{fig:LASSOGLOBAL}
\end{figure}

This fit includes the data of ETMC, HSC, CLS, [Prelovsek et al.] and RQCD given in Tables \ref{tab:ETMCJPsi}, \ref{tab:ETMCetac}, \ref{tab:hsc}, \ref{tab:prel} and \ref{tab:rqcd}. The data of PACS-CS are not included because of the tensions observed with the other data sets. In this case the information criteria, shown in Fig.~\ref{fig:LASSOGLOBAL} (top), have two minima, for $\lambda\simeq 0.5$ (global minima) and $\lambda\simeq 2.5$. This corresponds again to leaving out one parameter, $b$, or three, $b$, $d$ and $\Delta_\sigma$, see Fig.~\ref{fig:LASSOGLOBAL} (bottom). The results of the parameters before and after the LASSO are given in Tables~\ref{tab:pars} and \ref{tab:GLOBALpars}. 

\begin{table}[htb]
\setlength{\tabcolsep}{0.4em}
{\renewcommand{\arraystretch}{1.6}
\centering
\begin{tabular}{|c|c|c|c|c|c|} \hline
\multicolumn{2}{|c|}{Collaboration} & \multicolumn{2}{|c|}{$m_H$ (MeV)} & \multicolumn{2}{|c|}{$\Delta$ (MeV)} \\ \hline
Label & $a$ (fm) & (I) & (II) & (I) & (II) \\ \hline
\multirow{3}{*}{ETMC1} & $0.0619$  & $1937 \pm 4$ & $1945 \pm 4$ & $140 \pm 7$ & $141 \pm 7$ \\ \cline{2-6}
& $0.0815$ & $1949 \pm 5$ & $1956 \pm 5$ & $148 \pm 14$ & $148 \pm 14$ \\ \cline{2-6}
& $0.0885$ & $1955 \pm 5$ & $1963 \pm 5$ & $152 \pm 12$ & $153 \pm 12$ \\ \hline
\multirow{3}{*}{ETMC2} & $0.0619$ & $1950 \pm 3$ & $1958 \pm 3$ & $139 \pm 7$ & $140 \pm 7$ \\ \cline{2-6}
& $0.0815$ & $1973 \pm 5$ & $1980 \pm 5$ & $145 \pm 13$ & $146 \pm 13$ \\ \cline{2-6}
& $0.0885$ & $1988 \pm 4$ & $1996 \pm 4$ & $148 \pm 13$ & $149 \pm 13$ \\ \hline
ETMC & 0 & $1916$ & $1925$ & $131$ & $132$ \\ \hline
\multirow{2}{*}{HSC} & ${1/6079}^*$  & $1903 \pm 5$ & $1910 \pm 5$ & $99 \pm 7$ & $100 \pm 7$ \\ \cline{2-6}
& ${1/5667}^*$ & $1874 \pm 5$ & $1881 \pm 5$ & $91 \pm 8$ & $92 \pm 8$ \\ \hline
CLS & $0.08636$ & $1940 \pm 1$ & $1945 \pm 1$ & $95 \pm 2$ & $96 \pm 2$ \\ \hline
Prelovsek & $0.1239$ & $1574 \pm 2$ & $1582 \pm 2$ & $97 \pm 2$ & $98 \pm 2$ \\ \hline
RQCD & $0.0714$ & $1910 \pm 1$ & $1917 \pm 1$ & $98 \pm 1$ & $98 \pm 1$ \\ \hline
MILC & $0$ & $1913 \pm 4$ & $ 1920 \pm 4$ & $111 \pm 4$ & $112 \pm 4$\\ \hline
% $a_t=1/6079$ $a_t=1/5667$ GeV$^{-1}$
\end{tabular}}
\caption{Tree level parameters of the global fit. The lattice spacing with a $^*$, corresponding to HSC, are the temporal lattice spacing and they are shown in units of GeV$^{-1}$. The remaining parameters are shown in the first two rows of Table~\ref{tab:pars}.}
\label{tab:GLOBALpars}
\end{table}

\begin{figure}[h!]
   \begin{minipage}{0.453\textwidth}
     \centering
     \includegraphics[width=1.0\linewidth]{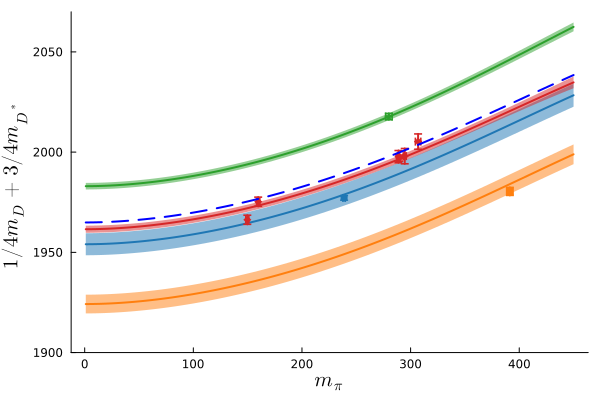}
   \end{minipage}
   \begin{minipage}{0.453\textwidth}
     \centering
     \includegraphics[width=1.0\linewidth]{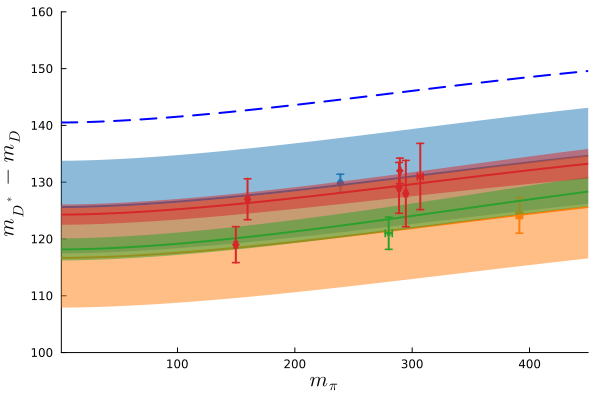}
   \end{minipage} 
   \begin{minipage}{0.453\textwidth}
     \centering
     \includegraphics[width=1.0\linewidth]{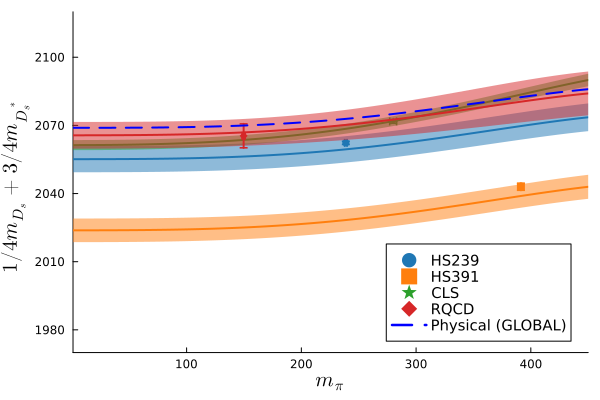}
   \end{minipage}
   \begin{minipage}{0.453\textwidth}
     \centering
     \includegraphics[width=1.0\linewidth]{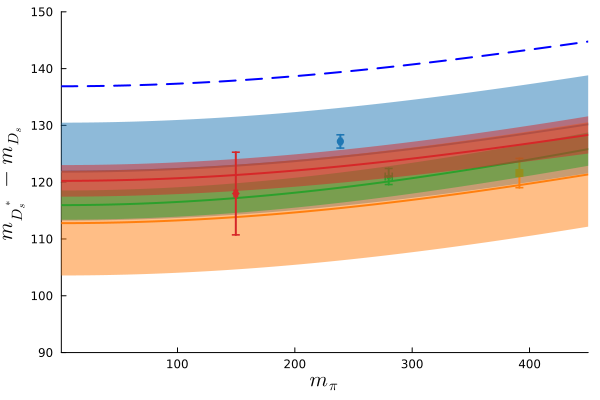}
   \end{minipage}
   \caption{Results of the $D$, $D^*$, $D_s$ and $D^*_s$ meson masses for the global analysis of some collaborations.}\label{fig:massesSINETMC}
\end{figure}

\begin{figure}[h!]
   \begin{minipage}{0.461\textwidth}
     \centering
     \includegraphics[width=1.0\linewidth]{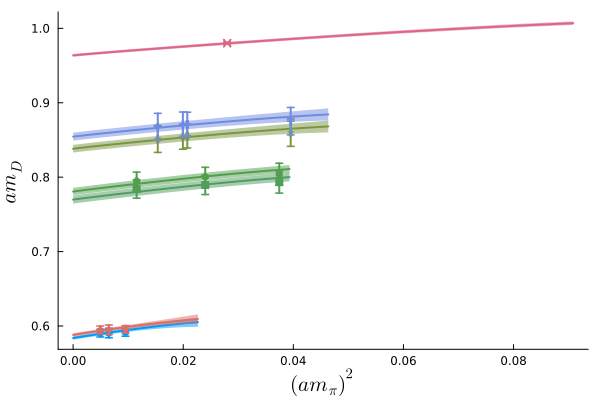}
   \end{minipage}
   \begin{minipage}{0.461\textwidth}
     \centering
     \includegraphics[width=1.0\linewidth]{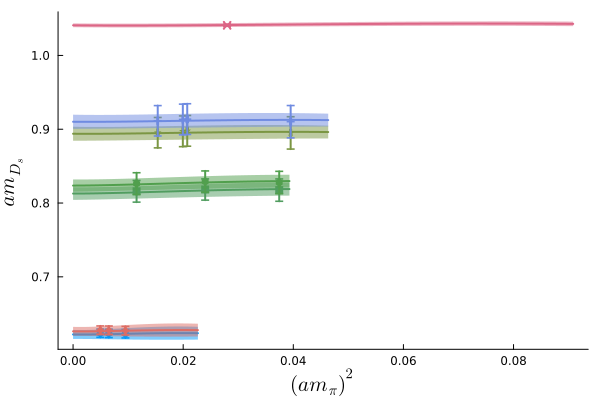}
   \end{minipage} 
   \begin{minipage}{0.461\textwidth}
     \centering
     \includegraphics[width=1.0\linewidth]{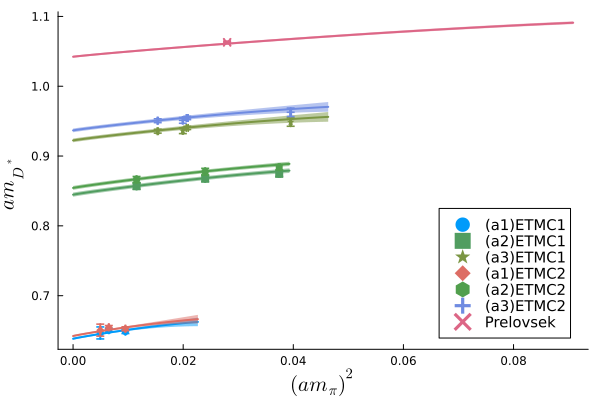}
   \end{minipage}
   \begin{minipage}{0.461\textwidth}
     \centering
     \includegraphics[width=1.0\linewidth]{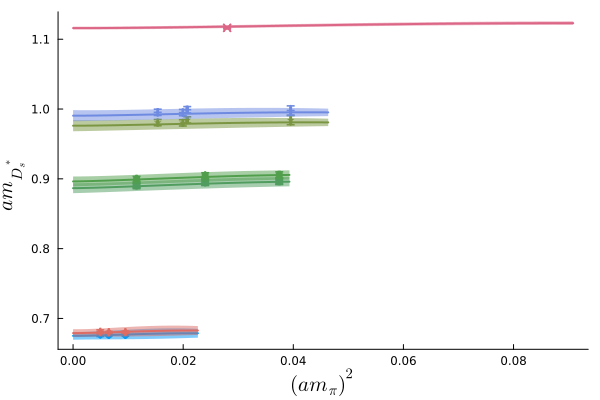}
   \end{minipage}
   \caption{As Fig.~\ref{fig:massesSINETMC} but with the remaining collaborations.}
   \label{fig:massesGLOBALETMC}
\end{figure}

In Figs.~\ref{fig:massesSINETMC} and \ref{fig:massesGLOBALETMC} we plot the results for the charmed meson masses in comparison with the data. In Fig.~\ref{fig:massesSINETMC} we plot the spin averaged masses and the hyperfine splittings of the charmed mesons as a function of the pion mass in physical units. We leave out ETMC and [Prelovsek et al.] data from this plot for a better view. In general, the error bands for ETMC data overlap with those from HSC, and the band for [Prelovsek et al.] data results in much smaller charmed masses. We show the results for these two data sets separately in Fig.~\ref{fig:massesGLOBALETMC} in units of the lattice spacing. As can be seen all data sets included in the global fit are very well described. The $\chi^2_{\mathrm{dof}}$ obtained is $0.85$. In Fig.~\ref{fig:massesSINETMC} we also show the result for the global fit fixing the $m_H$ and $\Delta$ parameters according to the experimental data. As can be seen, the \textit{physical} trajectories for the spin averaged charmed meson masses are closer to the ones of RQCD data, which includes an almost physical pion mass. Still, the hyperfine splittings of the charmed mesons in the lattice data are smaller than the \textit{physical} ones in $\sim8-15\%$. 
%, and the physical extrapolation, in comparison with the continuous extrapolations of the ETMC and MILC data.

We also performed an analysis with the tree level expressions, Eqs.~(\ref{eq:t1}) and ~(\ref{eq:t2}), and the results are given in Appendix \ref{app:tree}. The $\chi^2_\mathrm{dof}$ obtained is a bit larger than with the one-loop expressions, $1.11$, but it also admits a good fit. In general, the values of $m_H$ and $\Delta$ are larger than the ones obtained with the one-loop expressions, as expected. The values obtained for $\Delta$ are more reasonable for the one-loop expressions than for the tree level expressions, as can be seen comparing Table~\ref{tab:GLOBALpars} and Table~\ref{tab:tree1}, where for ETMC, larger values of $\Delta$ than the physical hyperfine splitting, are obtained using the tree level expressions.

\begin{figure}[h!]
   \begin{minipage}{0.45\textwidth}
     \centering
     \includegraphics[width=1.0\linewidth]{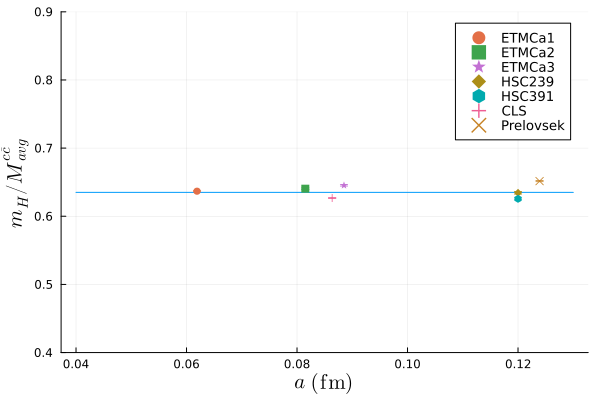}
   \end{minipage}
   \begin{minipage}{0.45\textwidth}
     \centering
     \includegraphics[width=1.0\linewidth]{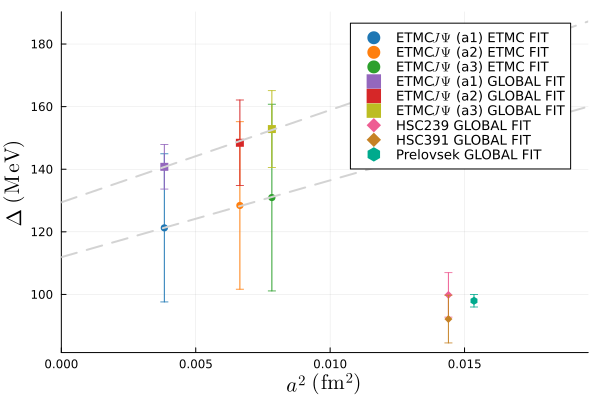}
   \end{minipage} 
   \caption{$m_H$ and $\Delta$ parameters as a function of the lattice spacing.}
   \label{fig:parcompa}
\end{figure}

It is interesting to see how the $m_H/M_{\mathrm{av}}^{c\bar{c}}$ and $\Delta$ parameters, with $ M_\mathrm{av}^{c\bar{c}}=(m_{\eta_c}+3m_{J/\psi})/4$, $m_H$ and $\Delta$ being the spin average mass and hyperfine splitting in the chiral limit, depend on the lattice spacing. This is shown in Fig.~\ref{fig:parcompa}. 
From Fig.~\ref{fig:parcompa}, one obtains that $m_H\simeq 0.63 M_\mathrm{av}^{c\bar{c}}$, where we expect this relation to be valid for charm quark masses beyond the physical one, but not very far from the range of the charmed quark masses of the LQCD simulations studied here\footnote{At the physical point, this relation deviates $\sim 10\%$ for the bottom counterparts assuming that the physical masses behave as leading order.}. Since $\Delta$ is a heavy quark spin violating parameter, one does not expect it to depend heavily on $m_H$. In Fig.~\ref{fig:parcompa}, taking into account the dependence of this parameter with the lattice spacing for the ETMC data, one sees that when $a\to 0$ this would lead to a range, $\Delta\sim 90-130$~MeV. In the future, more precise LQCD data can help to determine more precisely this parameter.

\section{Conclusions}
We have investigated the light and heavy quark mass dependence of the $D^{(*)}_{(s)}$ mesons by means of an analysis of the available lattice data from different LQCD Collaborations withing one-loop HH$\chi$PT. Model selection tools were implemented to determine the low energy constants with a higher precision. 

We have found that, concerning the higher order parameters, $b,c,d,\Delta_\sigma$ and $\Delta_a$, while $c$ and $\Delta_a$ are required by the precision of the LQCD data, $b,d$ and $\Delta_\sigma$ are less relevant, and, in general, LQCD data do not restrict the $b$ parameter. We have found more precise values of the one-loop parameters, once we have applied these model selection techniques. 

We also observe that, the largest difference in the various simulations comes mainly from the setting of the charm quark mass to the {\it physical point}, and that this difference becomes also larger due to discretization effects. Improvement on the methods to set the physical charm quark mass in the LQCD simulations are required at present, according to the present analysis. However, once these differences are taken into account, most of the LQCD data (all except for PACS) can be very well described together by one-loop HH$\chi$PT, considering different values of the tree level parameters, $m_H$ and $\Delta$. The data can be described well also with the tree level expressions taking into account one insertion of the light quark matrix instead of two, however, too high values of $\Delta$ are obtained in some cases.

The description with one loop involving pseudoscalar Goldstone bosons describes better the LQCD data and provides more reasonable values of the tree level parameters after the LASSO method is implemented. Once the charm quark mass is fixed within the theory as required by the experimental data, the charmed meson masses obtained from the global fit of LQCD data are consistent with the experimental data. 

\section{Acknowledgments}

%\begin{align}
%    \lim_{m\rightarrow 0} \frac{\partial M_{avg}}{\partial m} &=4\sigma +2a\delta_{1,2a} \\
%    \lim_{m\rightarrow 0} \frac{\partial Split.}{\partial m} &=4\Delta^{(\sigma)} +2\Delta^{(a)}\delta_{1,2a} \\
 %   \lim_{m\rightarrow 0} \frac{\partial m_D}{\partial m} &=4\sigma +2a -3\Delta^{(\sigma)} -\frac{3}{2}\Delta^{(a)} \\
  %  \lim_{m\rightarrow 0} \frac{\partial m_{D_s}}{\partial m} &=4\sigma  -3\Delta^{(\sigma)} \\
  %  \lim_{m\rightarrow 0} \frac{\partial m_{D^*}}{\partial m} &=4\sigma +2a +\Delta^{(\sigma)} +\frac{1}{2}\Delta^{(a)} \\
  %  \lim_{m\rightarrow 0} \frac{\partial m_{D_s^*}}{\partial m} &=4\sigma +\Delta^{(\sigma)}
%\end{align}

R. M. and F. Gil thank J. Komijani for providing the MILC continuous extrapolation to the physical point of the $D_{(s)}$ meson. We also thank A. Ramos and F. Romero for useful discussions. We  acknowledge support from the CIDEGENT program with Ref. CIDEGENT/2019/015 and from the spanish national grants PID2019-106080GB-C21 and PID2020-112777GB-I00. This project has received funding from the European Union’s Horizon 2020 programme No. 824093 for the STRONG-2020 project.

\clearpage
\appendix

\section{Coefficients in Eqs.~(\ref{eq:md1}) and ~(\ref{eq:md2})}
The coefficients $\alpha_l$, $\beta_l^{(X)}$, $(\gamma_l^{(X)}-\lambda_l^{(X)}\alpha_l)$, $c_l$, $(\gamma_l^{(X)}-\lambda_l^{(X)}\Delta)$ and $\delta c_l$, are given by,
%\begin{widetext}
\begin{align}
 \alpha_l=2\sigma(2m+m_s)+2a(m\delta_{1,2l}+m_s\delta_{3l})\end{align}\begin{align} \beta^{(\pi)}=g^2\left({\begin{array}{c}\frac{3}{2}\\\frac{3}{2}\\0\end{array}}\right);\,\beta^{(K)}=g^2\left({\begin{array}{c}1\\1\\2\end{array}}\right);\beta^{(\eta)}=g^2\left({\begin{array}{c}\frac{1}{6}\\\frac{1}{6}\\\frac{2}{3}\end{array}}\right)\label{eq:par1}
 \end{align}
\begin{align}(\gamma_l^{(\pi)}-\lambda_l^{(\pi)}\alpha_l)=&-a(3m\delta_{1l}+\delta_{2l})-6\sigma m\nonumber\\
(\gamma_l^{(K)}-\lambda_l^{(K)}\alpha_l)=&-a(m+m_s)(\delta_{1l}+\delta_{2l})-2\sigma(m+m_s) \nonumber\\
&+3g^2a(m_s-m)(\delta_{1l}-\delta_{3l})\nonumber\\
(\gamma_l^{(\eta)}-\lambda_l^{(\eta)}\alpha_l)=&-a\left\{\frac{1}{3}m(\delta_{1l}+\delta_{2l})+\frac{4}{3}m_s\delta_{3l}\right\} \nonumber\\
&-\frac{\sigma}{3}(2m+4m_s)\nonumber\\
(\gamma_l^{(X)}-\lambda_l^{(X)}\Delta)=&-\Delta\frac{2}{3}\lambda^{(X)}_l=-\Delta\beta^{(X)}_l\label{eq:par2}
\end{align}
\begin{align}
 c_l=&\,4b(m^2\delta_{1,2l}+m_s^2\delta_{3l})+4c(2m+m_s)(m\delta_{1,2l}+m_s\delta_{3l}) \nonumber\\
 &+4d(2m^2+m_s^2)\nonumber\\
 \delta c_l=&\,2\Delta^{(\sigma)}(2m+m_s)+2\Delta^{(a)}(m\delta_{1,2l}+m_s\delta_{3l})\,\label{eq:par3}
\end{align}
%\end{widetext}

where we have assumed the isospin limit $m=m_u=m_d$ and $\delta_{1,2l}=\delta_{1,l}+\delta_{2,l}$. For the general formulas taking into account isospin breaking see Ref.~\cite{jenkins}. The light quark masses can be related to the leading order pseudoscalar masses, $M_{X}^0$, like
\begin{eqnarray}
m=\frac{M^{02}_{\pi}}{2B_0}, \qquad m_s=\frac{M^{02}_{K}}{B_0}-\frac{M^{02}_{\pi}}{2B_0}.
\end{eqnarray}
For simplicity, we identify the physical masses of the light pseudoscalar meson masses with the leading order pseudoscalar masses, $M_\pi=M^0_\pi$ and $M_K=M^0_K$, $M_\eta^2=M_\eta^{02}=(4M_K^{02}-M_\pi^{02})/3$.

\section{Results for the tree level fits of the charmed meson masses}\label{app:tree}

In this section we present the results of analyzing the charmed meson masses from LQCD with the tree level expressions including one insertion of the light quark mass matrix. The relations in these case are,
%\begin{widetext}
\begin{align}
   \frac{1}{4}(M_{P_l}&+3M_{P_l^*})= \nonumber\\
   & m_H+2\sigma(2m+m_s)+2a(m\delta_{1,2l}+m_s\delta_{3l})\label{eq:t1}\ ,
\end{align}  
\begin{align}
   M_{P_l}^*&-M_{P_l}=\nonumber\\
   &\Delta +2\Delta^{(\sigma)}(2m+m_s)+2\Delta^{(a)}(m\delta_{1,2l}+m_s\delta_{3l})\ .\label{eq:t2}
\end{align}    
%\end{widetext}
The parameters obtained in the various analyses using these relations are given in Tables \ref{tab:tree1} and \ref{tab:tree2}. The $\chi^2_{dof}$ obtained is also good, $0.42$ for the ETMC fit and $1.11$ for the Global fit. For MILC we obtain a value close to zero because of the refitting of the continuously extrapolated function.

\begin{table}[h!]
\setlength{\tabcolsep}{0.4em}
{\renewcommand{\arraystretch}{1.6}
\centering
\begin{tabular}{|c|c|c|c|c|} \hline
Fit & $\frac{\sigma m_\pi}{{B_0}}$\scriptsize{$\cdot 10^{2}$} & $\frac{a m_\pi}{{B_0}}$\scriptsize{$\cdot 10^{2}$} & $\frac{\Delta^{(\sigma)}m_\pi}{{B_0}}$\scriptsize{$\cdot 10^{3}$}& $\frac{\Delta^{(a)}m_\pi}{{B_0}}$\scriptsize{$\cdot 10^{3}$} \\ \hline
\scriptsize{ETMC} & $0.22 \pm 0.23$ & $2.80 \pm 0.10$ & $8.58 \pm 6.60$ & $-2.79 \pm 3.09$  \\ \hline
\scriptsize{MILC} & $1.15 \pm 0.36$ & $3.38 \pm 0.51$ & $8.38 \pm 2.42$ & $5.18 \pm 6.17$  \\ \hline
\scriptsize{GLOBAL} & $0.95 \pm 0.01$ & $2.73 \pm 0.04$ & $0.66 \pm 0.32$ & $-2.59 \pm 0.65$ \\ \hline
\end{tabular}}
\caption{Results of the analyses with the tree level formulas, Eqs.~(\ref{eq:t1}) and (\ref{eq:t2}). The remaining parameters are shown in Table~\ref{tab:tree1}}
\label{tab:tree2}
\end{table}

\begin{table}[htb!]
\setlength{\tabcolsep}{0.4em}
{\renewcommand{\arraystretch}{1.6}
\centering
\begin{tabular}{|c|c|c|c|c|} \hline
Fit & Label & $a$ (fm) & $m_H$ (MeV) & $\Delta$ (MeV) \\ \hline
\multirow{7}{*}{ETMC} & \multirow{3}{*}{ETMC1} & $0.0619$  & $1994 \pm 12$ & $137 \pm 34$ \\ \cline{3-5}
& & $0.0815$ & $2010 \pm 14$ & $145 \pm 39$ \\ \cline{3-5}
& & $0.0885$ & $2014 \pm 14$ & $147 \pm 42$ \\ \cline{2-5}
& \multirow{3}{*}{ETMC2} & $0.0619$  & $2006 \pm 12$ & $138 \pm 33$ \\ \cline{3-5}
& & $0.0815$ & $2034 \pm 13$ & $143 \pm 39$ \\ \cline{3-5}
& & $0.0885$ & $2047 \pm 14$ & $146 \pm 39$ \\ \cline{2-5}
& ETMC  & $0$ & $1971$ & $129$ \\ \hline
MILC & MILC & $0$ & $1952 \pm 21$ & $140$ \\ \hline
\multirow{12}{*}{GLOBAL} & \multirow{3}{*}{ETMC1} & $0.0619$  & $1959 \pm 3$ & $178 \pm 8$ \\ \cline{3-5}
& & $0.0815$ & $1972 \pm 5$ & $188 \pm 16$ \\ \cline{3-5}
& & $0.0885$ & $1974 \pm 4$ & $194 \pm 14$ \\ \cline{2-5}
& \multirow{3}{*}{ETMC2} & $0.0619$  & $1971 \pm 3$ & $177 \pm 8$ \\ \cline{3-5}
& & $0.0815$ & $1996 \pm 5$ & $185 \pm 14$ \\ \cline{3-5}
& & $0.0885$ & $2007 \pm 4$ & $190 \pm 15$ \\ \cline{2-5}
& \multirow{2}{*}{HSC} & $1/6079$& $1929 \pm 5$ & $129 \pm 8$ \\ \cline{3-5}
& & $1/5667$ & $1899 \pm 5$ & $124 \pm 9$ \\ \cline{2-5}
& CLS & $0.08636$ & $1966 \pm 1$ & $124 \pm 2$ \\ \cline{2-5}
& Prelovsek & $0.1239$ & $1596 \pm 1$ & $128 \pm 3$ \\ \cline{2-5}
& RQCD & $0.0714$ & $1934 \pm 1$ & $127 \pm 2$ \\ \cline{2-5}
& MILC & $0$ & $1938 \pm 3$ & $144 \pm 5$ \\ \hline
\end{tabular}}
\caption{Results of the analyses with the tree level formulas, Eqs.~(\ref{eq:t1}) and (\ref{eq:t2}). The remaining parameters are shown in Table~\ref{tab:tree2}.}
\label{tab:tree1}
\end{table}

\newpage
\begin{widetext}
\section{Data included in the fits of \ref{sec:results}}\label{app:data}
%\end{widetext}
%\begin{widetext}

\begin{table*}[htb!]
\setlength{\tabcolsep}{0.8em}
{\renewcommand{\arraystretch}{1.6}
\centering
\begin{tabular}{|c|c|c|c|c|c|c|c|c|} \hline
$a$ (fm) & $(L/a)^3\times T/a$ & $am_\pi$ & $am_K$ & $am_D$ & $am_{D_s}$ & $am_{D^*}$ & $am_{D_s^*}$ & $am_{J/\Psi}$\\ \hline
    \multirow{3}{*}{$0.0619(18)$} & $48^3\times 96$ & 0.0703 & 0.1697 & 0.5905 & 0.6236 & 0.6466 & 0.6770 & 0.9715 \\ \cline{2-9}
    & $48^3\times 96$ & 0.0806 & 0.1738 & 0.5906 & 0.6234 & 0.6506 & 0.6763 & 0.9697 \\ \cline{2-9}
    & $48^3\times 96$ & 0.0975 & 0.1768 & 0.5913 & 0.6229 & 0.6486 & 0.6764 & 0.9703 \\ \hline
    \multirow{3}{*}{$0.0815(30)$} & $32^3\times 64$ & 0.1074 & 0.2133 & 0.7840 & 0.8159 & 0.8568 & 0.8905 & 1.2791 \\ \cline{2-9}
    & $32^3\times 64$ & 0.1549 & 0.2279 & 0.7895 & 0.8183 & 0.8678 & 0.8950 & 1.2828 \\ \cline{2-9}
    & $24^3\times 48$ & 0.1935 & 0.2430 & 0.7934 & 0.8175 & 0.8745 & 0.8965 & 1.2818 \\ \hline
    \multirow{3}{*}{$0.0885(36)$} & $32^3\times 64$ & 0.1240 & 0.2512 & 0.8514 & 0.8953 & 0.9356 & 0.9806 & 1.3890 \\ \cline{2-9}
    & $32^3\times 64$ & 0.1412 & 0.2569 & 0.8544 & 0.8972 & 0.9363 & 0.9802 & 1.3895 \\ \cline{2-9}
    & $24^3\times 48$ & 0.1440 & 0.2589 & 0.8552 & 0.8978 & 0.9403 & 0.9844 & 1.3906 \\ \cline{2-9}
    & $24^3\times 48$ & 0.1988 & 0.2764 & 0.8599 & 0.8950 & 0.9487 & 0.9841 & 1.3882 \\ \hline
\end{tabular}}
\caption{$D^{(*)}_{(s)}$ and $J/\Psi$ meson masses. Data collected from \cite{guoheo} provided by the authors of \cite{kalinowskiwagner}. The charm quark mass is fixed by reproducing the physical $J/\psi$ meson mass. Lattice spacing uncertainties are discused on \cite{EuropeanTwistedMassa}. $N_F=2+1+1$.}
\label{tab:ETMCJPsi}
\end{table*}

\begin{table*}[htb!]
 \setlength{\tabcolsep}{0.8em}
{\renewcommand{\arraystretch}{1.6}
\centering
\begin{tabular}{|c|c|c|c|c|c|c|c|c|c|} \hline
 $a$ (fm) & $(L/a)^3\times T/a$ & $am_\pi$ & $am_K$ & $am_D$ & $am_{D_s}$ & $am_{D^*}$ & $am_{D_s^*}$ & $am_{\eta_c}$ \\ \hline
    \multirow{3}{*}{$0.0619(18)$} & $48^3\times 96$ & 0.0703 & 0.1697 & 0.5947 & 0.6279 & 0.6506 & 0.6809 & 0.9351 \\ \cline{2-9}
    & $48^3\times 96$ & 0.0806 & 0.1738 & 0.5949 & 0.6277 & 0.6546 & 0.6803 & 0.9332 \\ \cline{2-9}
    & $48^3\times 96$ & 0.0975 & 0.1768 & 0.5955 & 0.6271 & 0.6526 & 0.6804 & 0.9335 \\ \hline
    \multirow{3}{*}{$0.0815(30)$} & $32^3\times 64$ & 0.1074 & 0.2133 & 0.7946 & 0.8263 & 0.8664 & 0.9001 & 1.2312 \\ \cline{2-9}
    & $32^3\times 64$ & 0.1549 & 0.2279 & 0.8004 & 0.8291 & 0.8777 & 0.9049 & 1.2342 \\ \cline{2-9}
    & $24^3\times 48$ & 0.1935 & 0.2430 & 0.8039 & 0.8278 & 0.8840 & 0.9059 & 1.2314 \\ \hline
    \multirow{3}{*}{$0.0885(36)$} & $32^3\times 64$ & 0.1240 & 0.2512 & 0.8677 & 0.9114 & 0.9506 & 0.9953 & 1.3370 \\ \cline{2-9}
    & $32^3\times 64$ & 0.1412 & 0.2569 & 0.8708 & 0.9132 & 0.9511 & 0.9949 & 1.3379 \\ \cline{2-9}
    & $24^3\times 48$ & 0.1440 & 0.2589 & 0.8714 & 0.9137 & 0.9545 & 0.9990 & 1.3382 \\ \cline{2-9}
    & $24^3\times 48$ & 0.1988 & 0.2764 & 0.8753 & 0.9102 & 0.9627 & 0.9980 & 1.3325 \\ \hline
\end{tabular}}
\caption{The same as Table \ref{tab:ETMCJPsi} but with $D^{(*)}_{(s)}$ and $\eta_c$ meson masses. The charm quark mass is
fixed by reproducing the physical $\eta_c$ meson mass.}
\label{tab:ETMCetac}
\end{table*}

\begin{table*}[htb!]
 \setlength{\tabcolsep}{0.8em}
{\renewcommand{\arraystretch}{1.6}
\centering
\begin{tabular}{|c|c|c|c|c|c|c|c|c|} \hline
$a$ (fm) & $am_\pi$ & $am_K$ & $am_D$ & $am_{D_s}$ & $am_{D^*}$ & $am_{D_s^*}$ & $am_{\eta_c}$ & $am_{J/\Psi}$ \\ \hline
    \multirow{3}{*}{$a =0.0907(13)$}& 0.32242 & 0.36269 & 0.84022 & 0.86273 & 0.91237 & 0.93297 & 1.22924 & 1.27877 \\ \cline{2-9}
    & 0.26191 & 0.32785 & - & 0.85158 & - & 0.91864 & 1.22098 & 1.26860 \\ \cline{2-9}
    & 0.18903 & 0.29190 & 0.79580 & 0.84000 & 0.86327 & 0.90429 & 1.21369 & 1.25988 \\ \cline{2-9}
    & 0.13593 & 0.27282 & 0.78798 & 0.83929 & 0.85776 & 0.90429 & 1.21268 & 1.25904 \\ \cline{2-9}
    & 0.07162 & 0.25454 & 0.77646 & 0.83149 & 0.83656 & 0.89268 & 1.20765 & 1.25258 \\ \hline
\end{tabular}}
\caption{Data for heavy meson masses from \cite{mohlerwoloshyn} and light meson masses collected from \cite{aokiphys} [PACS-CS]. The charm quark hopping pa
rameter $\kappa_c$ has been tuned to the value where the spin-averaged kinetic mass $(M_{D_s} +3M_{D_{s}^*} )/4$ takes its physical value. The lowest pion mass data is also named as Ensemble 2 in \cite{langmohler}. Also, $(L/a)^3\times T/a=32^3\times 64$ and $N_F=2+1$.}
\label{tab:pacs}
\end{table*}

\newpage
\begin{table*}[htb!]
\setlength{\tabcolsep}{0.8em}
{\renewcommand{\arraystretch}{1.6}
\centering
\begin{tabular}{|c|c|c|c|c|c|c|c|} \hline
$a_t^{-1}$ (MeV) & $(L/a_s )^3\times (T /a_t )$ & $m_\pi$ & $m_K$ & $m_D$ & $m_{D_s}$ & $m_{D^*}$ & $m_{D_s^*}$ \\ \hline
    6079 & $32^3\times 256$ & 0.03928(18) & 0.08344(7) & 0.30923(11) & 0.32356(12) & 0.33058(24) & 0.34448(15)     \\ \hline
  5667 & $(16,20,24)^3\times 128$ & 0.06906(13) & 0.09698(9) & 0.33303(31) & 0.34441(29) & 0.35494(46) & 0.36587(35)  \\ \hline
\end{tabular}}
\caption{Data collected from \cite{cheungthomas} for $N_F=2+1$ [HSC]. The charm quark mass parameters have been determined using the $\eta_c$ physical mass, obtaining the values, $(a_t m_{\eta_c})^2=0.2412$ and $0.2735$ for the light and heavy pion mass respectively (see Figs. 1 of \cite{cheungohara,liuming}). %\textcolor{green}{DK scattering data}.} 
}
%16 3 × 128
%20 3 × 128
%24 3 × 128
\label{tab:hsc}
\end{table*}

\begin{table*}[htb!]
 \setlength{\tabcolsep}{0.8em}
{\renewcommand{\arraystretch}{1.6}
\centering
\begin{tabular}{|c|c|c|c|c|c|c|c|} \hline
 $a$ (fm) & $m_\pi$ & $m_K$ & $m_D$ & $m_{D_s}$ & $m_{D^*}$ & $m_{D_s^*}$ & $M_{avg}$ \\ \hline
    \multirow{1}{*}{$a =0.08636(138)$} & 280 & 467 & 1927 & 1981 & 2048 & 2102 & 3103 \\ \hline
\end{tabular}}
\caption{Data collected from \cite{prelovsekpadmanath} where the spin average $M_{avg}=(m_{\eta_c}+3m_{J/\Psi})/4$ has been tune to the physical point, for $(L/a)^3\times T/a=24^3\times 128$ and $N_F=2+1$ [CLS]. Masses given in MeV.}
\label{tab:cls}
\end{table*}

\begin{table*}[htb!]
 \setlength{\tabcolsep}{0.8em}
{\renewcommand{\arraystretch}{1.6}
\centering
\begin{tabular}{|c|c|c|c|c|c|c|c|} \hline
$a$ (fm) & $am_\pi$ & $am_K$ & $am_D$ & $am_{D_s}$ & $am_{D^*}$ & $am_{D_s^*}$ & $aM_{avg}$ \\ \hline
    \multirow{1}{*}{$a =0.1239(13)$} & 0.1673 & 0.3467 & 0.9801 & 1.04075433 & 1.0629 & 1.11635248 & 1.52499 \\ \hline
\end{tabular}}
\caption{Data collected from Ensemble 1 in \cite{langmohler} for $(L/a)^3\times T/a=16^3\times 32$ and $N_F=2$ [Prelovsek et al.]. The spin-average $(m_{\eta_c}+3m_{J/\psi})/4$ is tuned to reproduce its physical value. %\textcolor{green}{DK scattering data together with last row of PACS}.
}
\label{tab:prel}
\end{table*}

\begin{table*}[htb!]
  \setlength{\tabcolsep}{0.8em}
{\renewcommand{\arraystretch}{1.6}
\centering
\begin{tabular}{|c|c|c|c|c|c|c|c|} \hline
 $a$ (fm) & $(L/a)^3\times T/a$ & $m_\pi$ & $m_K$ & $m_D$ & $m_{D_s}$ & $m_{D^*}$ & $m_{D_s^*}$ \\ \hline
    \multirow{6}{*}{$a =0.071$} & $24^3\times 48$ & 306.9 & 540 & 1907 & - & 2038 & - \\ \cline{2-8}
    & $32^3\times 64$ & 294.6 & 528 & 1902 & - & 2030 & - \\ \cline{2-8}
    & $40^3\times 64$ & 288.8 & 527 & 1901 & - & 2030 & - \\ \cline{2-8}
    & $64^3\times 64$ & 289.5 & 526 & 1898 & - & 2030 & - \\ \cline{2-8}
    & $48^3\times 64$ & 159.7 & 500 & 1880 & - & 2007 & - \\ \cline{2-8}
    & $64^3\times 64$ & 149.7 & 497 & 1877 & 1976.9 & 1996 & 2094.9 \\ \hline
\end{tabular}}
\caption{Data collected from table I in \cite{balicollins} for $N_F=2$ [RQCD]. The charm quark mass is fixed through the experimental value of the spin-averaged
$1S$ charmonium mass, $m_{1S} = 3068.5 $ MeV. Masses given in MeV. %\textcolor{green}{DK scattering data}.
}
\label{tab:rqcd}
\end{table*}
\end{widetext}
%\end{widetext}

\clearpage

\bibliography{biblio}

\end{document}